\documentclass[preprintnumbers,amsmath,11pt,amssymb,floatfix,superscriptaddress,nofootinbib]{revtex4}
\usepackage{graphicx}
\usepackage{epsfig}
\usepackage{bm}
\usepackage{amsfonts}
\renewcommand\thesection{\arabic{section}}

\begin{document}
\title{\Large A look into the cosmological consequences of a dark energy model with higher derivatives of $H$ in the framework of Chameleon Brans-Dicke Cosmology }
\author{Antonio Pasqua}
\email{toto.pasqua@gmail.com} \affiliation{Department of Physics,
University of Trieste, Via Valerio, 2 34127 Trieste, Italy.}

\author{Surajit Chattopadhyay}
\email{schattopadhyay1@kol.amity.edu; surajitchatto@outlook.com}
\affiliation{ Department of Mathematics, Amity University, Major
Arterial Road, Action Area II, Rajarhat, New Town, Kolkata 700135,
India.}

\author{Aroonkumar Beesham}
\email{beeshama@unizulu.ac.za} \affiliation{ Department of
Mathematical Sciences, University of Zululand Private Bag X1001,
Kwa-Dlangezwa 3886, South Africa.}

\date{\today}

\begin{abstract}
\textbf{Abstract}: In this paper, we study some relevant
cosmological features of a Dark Energy (DE) model with
Granda-Oliveiros cut-off, which is just a specific case of
Nojiri-Odintsov holographic DE (ref \cite{odi1}) unifying phantom
inflation with late-time acceleration, in the framework of
Chameleon Brans-Dicke (BD) Cosmology. Choosing a particular ansatz
for some of the quantities involved, we derive the expressions of
some important cosmological quantities, like the Equation of State
(EoS) parameter of DE $w_D$, the effective EoS parameter
$w_{eff}$, the pressure of DE $p_D$ and the deceleration parameter
$q$. Moreover, we study the behavior of statefinder parameters $r$
and $s$, of the cosmographic parameters $j$, $s_{cosmo}$, $l$ and
$m$ and of the squared speed of the sound $v_s^2$ for both case
corresponding to non interacting and interacting Dark Sectors. We
also plot the quantities we have derived and we calculate their
values for $t\rightarrow 0$ (i.e. for the beginning of the
Universe history), for $t\rightarrow \infty$ (i.e. for far future)
and for the present time, indicated with $t_0$. The EoS parameters
have been tested against various observational values available in
the literature.
\end{abstract}

\maketitle

\section{Introduction}
Accumulating the observational data of distant Supernovae Ia,
Riess et al. \cite{1b}  and Perlmutter et al. \cite{1a}
independently reported that the current Universe is expanding with
acceleration. Subsequently, other observational studies including
Cosmic Microwave Background (CMB) radiation anisotropies, Large
Scale Structure (LSS) and X-ray experiments have  provided further
strong evidences supporting the present day accelerated expansion
of the Universe \cite{1c,1d,1h,cmb2,cmb3,planck,sds1,sds2,xray}.
The exotic matter, perhaps the cause of this acceleration, is
dubbed as ``Dark Energy" (DE).  DE is different from the ordinary
matter in the sense
 that it is characterized by negative pressure. The The Cosmological Constant $\Lambda_{CC}$,
 proposed first by Einstein and latter abandoned by himself, revived as the simplest candidate
 of DE once the accelerated expansion is discovered. There are two other approaches to account for this late time acceleration. One is to introduce a suitable energy-momentum tensor with negative pressure to the right hand side of Einstein equation and other is to modify the left hand side of Einstein equation. This first approach is dubbed as ``modified matter" approach and the one is called ``modified gravity" approach. Different candidates for late time acceleration are differentiated based on the EoS parameter $w=p/\rho$, which for accelerated expansion of the Universe requires to be $<-1/3$. The Cosmological Constat $\Lambda_{CC}$ has a constant EoS parameter which value is exactly $w=-1$. Although it is consistent with observations, it can not tell us much about the time evolution of the EoS parameter and hence many other DE models have been proposed where the EoS parameter is dynamic in nature.  These models are reviewed in \cite{copeland-2006,delcampo,delcampoa,delcampob,delcampoc,delcampod,delcampoe,delcampof,delcampog,delcampoh}.\\
According to recent observational data,  Dark Energy (DE) and Dark Matter (DM) represent,
respectively, the 68.3$\%$ and  the 26.8$\%$ of the total energy density $\rho_{tot}$ of the present day observable Universe \cite{twothirds},  while the ordinary baryonic matter contributes for only about the 4.9$\%$ of $\rho_{tot}$. Moreover, we have that the radiation term $\rho_r$ contributes to the total cosmic energy density $\rho_{tot}$ in a practically negligibly way. Different  candidates of DE have been discussed widely  in scientific literature: some of them include  quintessence, tachyon,  k-essence, Chaplygin gas, quintom, Agegraphic DE (ADE), New Agegraphic (NADE)  and phantom  \cite{dil1,dil2,dil2-1,kess1,kess3,kess4,kess2-2,quint1,quint3,
quint5,quint6,quint7,tac3,tac1-1,tac1-2,tac1-3,tac1-4,tac2-1,tac2-3,tac2-4,
pha1,pha3,pha4,pha5,pha6,pha7,qui1,qui2,qui3,qui4,qui5,qui6,qui8,qui10,qui12,cgas1,cgas2,cgas3,ade1,ade2,2,2a}.\\
In literature, one of the most studied DE candidate model
is represented by the so-called Holographic DE (HDE) model \cite{3,3b,4,5,5a,5b,6,7},
 which is based on the ``holographic principle", according to which the total entropy
 of a system scales not with the volume $V$, but instead with the area $A$ of the enveloping horizon.
 In the context of Cosmology, it sets an upper bound to the entropy of the Universe.
 It was shown by Cohen et al. \cite{7}  that in Quantum Field Theory (QFT) the ultraviolet (UV) cut-off $\Lambda_{UV}$ should be related to the IR cut-off $L$ due to limit set by forming a black hole. If the vacuum energy density caused by UV cut-off is given by the relation $\rho_D = \Lambda^4_{UV}$, then the total energy of a system of size $L$ should  be less or equal than the mass of the system-size black hole, i.e. we must have that:
\begin{eqnarray}
E_D \leq E_{BH}, \label{1old}
\end{eqnarray}
which implies that:
\begin{eqnarray}
L^3 \rho_D \leq M_p^2 L, \label{1}
\end{eqnarray}
where the term $M_p = \left( 8\pi G  \right)^{-1/2}  \approx 10^{18} $GeV represents the reduced Planck mass (with $G$ being the Newton's gravitational constant).  If the largest possible cut-off $L$ is the one which saturate the inequality given in Eq. (\ref{1}), we obtain the following relation for the energy density of HDE $\rho_D$:
\begin{eqnarray}
\rho_D = 3c^2 M_p^2 L^{-2},   \label{2}
\end{eqnarray}
where the quantity $c^2$ represents a dimensionless numerical constant which precise value  can be derived by  using observational cosmological data:  for  a flat (i.e. for $k=0$) Universe $c=0.818_{-0.097} ^{+0.113}$  and in the case of a non-flat (i.e. for $k\neq 0$) Universe we have  $c=0.815_{-0.139} ^{+0.179} $ \cite{n2primo,n2secondo}.\\
Recently, the cosmic acceleration has been also accurately studied by using the concept of modification of gravity  \cite{nojo,nojo2}. For example, we know that adding a term proportional to $1/R$ (with $R$ indicating the Ricci scalar) to the Einstein action leads to a gravitational alternative to the DE \cite{frieman1,frieman2}.\\
 Some of the most famous and known models of modified gravity are represented by braneworld models, $f\left(T\right)$ gravity (where the term $T$ indicates the torsion scalar),
$f \left(R\right)$ gravity (where the term $R$ indicates the Ricci scalar curvature), $f \left(G\right)$ gravity (where the term $G=R^2-4R_{\mu \nu}R^{\mu \nu} + R_{\mu \nu \lambda \sigma}R^{\mu \nu \lambda \sigma}$ represents the Gauss-Bonnet invariant, with $R_{\mu \nu}$ representing the Ricci curvature tensor and $R_{\mu \nu \lambda \sigma}$ representing the Riemann curvature tensor), $f \left(R,T\right)$ gravity, Horava-Lifshitz gravity Dirac-Born-Infeld (DBI) models,  Dvali-Gabadadze-Porrati (DGP) model and Brans-Dicke (BD) gravity \cite{15,15a,15b,15c,15e,15g,15i,15l,mio,miobd2,dgp1,miodbi,miohl,frt2,frt5,miofg1,fr1,fr2,fr3,miofr,fr7,fr8,fr10,fr11,fr12,fr13,fr14,fr15,
mioft1,mioft2,mioft3,ft1,ft2,ft3,ft5,ft6,ft8,bra1,bra2,miors,antonico1,antonico2,antonico3}.\\
In the present work, we focus on a recently proposed energy density model which is function of the Hubble parameter $H$ and on its first and second time derivatives in the framework of Chameleon Brans-Dicke Cosmology. Therefore, we here discuss the main characteristics and features of the BD Chameleon Cosmology. Cosmological models of the classical BD theory were first studied in the papers of Greenstein \cite{24,25}.
Later on, Khoury $\&$ Weltman \cite{34,35} considered self-interactions of the scalar field in order to avoid the bounds on such a field; moreover, they suggested such scalars to be Chameleon fields because of the way in which the mass of the fields depends on the energy density of matter in the local environment. In another work, Brax et al. \cite{36} have demonstrated that the  Chameleon scalar field can produce explicit realizations of a quintessence model, where the quintessence scalar field directly couples to baryons and DM with gravitational strength. Farajollahi $\&$ Salehi \cite{37} discussed the important role of interacting Chameleon scalar field in the phantom crossing. The interacting HDE model in the framework of the Chameleon-tachyon Cosmology was discussed in the paper of Farajollahi et al. \cite{38}.\\
A Chameleon scalar field is introduced in models with a non-minimal coupling between the scalar field and the matter system \cite{33}. Das $\&$ Banerjee \cite{40}  introduced this kind of Chameleon-matter coupling in the BD model to achieve an accelerated expansion of the Universe. In a recent work, Bisabr \cite{33} considered a generalized BD model with the scale factor $a\left(t\right)$ given in the power law form and allowed a non-minimal coupling with the matter sector.\\
Since DE occupies about 70$\%$ of the total energy density of the present day Universe, while sooner after the Big Bang its contribution was practically negligible, it is reasonable to consider that the energy density of DE must be a function of the Hubble parameter $H$ and its derivatives with respect to the cosmic time $t$ since the Hubble parameter gives information about the expansion rate of the Universe. In view of this, Chen $\&$ Jing \cite{modelhigher} recently assumed that the energy density $\rho_D$ of DE contains three terms, one proportional to the Hubble parameter $H$ and other two which are, respectively, proportional to the first and to the second time derivatives of $H$, i.e. $\dot{H}$ and $\ddot{H}$. Each term forming the expression of the energy density is also proportional to a constant parameter. The final expression of the energy density $\rho_D$ of DE they proposed is given by the following relation:
\begin{eqnarray}
\rho_D = 3M_p^2\left( \alpha \ddot{H}H^{-1}+\beta\dot{H}+\varepsilon H^2   \right), \label{model}
\end{eqnarray}
where $\alpha$, $\beta$ and $\varepsilon$ are three arbitrary dimensionless parameters characterizing the DE energy density model. One of the main properties of this model is that it can help to alleviate the age problem of the old objects. The numerical factor 3 present in Eq. (\ref{model}) is introduced since it helps to simplify some of the following calculations. We must emphasize here that the inverse of the Hubble parameter (i.e. $H^{-1}$) is introduced in the first term of the energy density of DE given in Eq. (\ref{model}) so that the dimensions of all terms are identical. The behavior and the main cosmological characteristics of the model given in Eq. (\ref{model}) strongly depend on the three parameters of the model $\alpha$, $\beta$ and $\varepsilon$. The expression of the energy density $\rho_D$ given in Eq. (\ref{model}) can be considered as an extension and a generalization of other DE energy density models previously proposed in scientific literature. In fact, in the limiting case corresponding to $\alpha = 0$, we obtain the energy density of DE in the case the infrared cut-off of the system is given by the Granda-Oliveros (GO) cut-off \cite{goli}. Moreover, in the limiting case corresponding to $\alpha = 0$, $\beta = 1$ and $\varepsilon = 2$, we obtain the expression of the energy density of DE with IR cut-off proportional to the average radius of the Ricci scalar curvature $\left(L \propto R^{-1/2}   \right)$ in the case of curvature parameter $k$ assumes the value of zero, i.e. $k=0$.  Since the model considered in this work has an extra free parameter,  it can be considered more general than the RDE model.\\
Other interesting works involving higher derivatives of the Hubble parameter are \cite{grass1,grass2}.\\
The paper is organized as follows. In Section 2, we write the main information about the cosmological properties of the Chameleon BD Cosmology, the energy density model we are considering and the ansatz we have chosen for some parameters considered in this paper; moreover, we obtain some important cosmological quantities, like the Equation of State (EoS) parameter of DE $w_D$, the effective EoS parameter $w_{eff}$, the pressure of DE $p_D$ and the deceleration parameter $q$. In Section 3, we study the statefinder diagnostic for the model we are studying. In Section 4, we derive and study the expressions of the Cosmographic parameters $j$, $s_{cosmo}$, $l$ and $m$ for the model we consider. In Section 5, we study the behavior of the squared speed of the sound $v_s^2$  in order to have more information about the model we are dealing with, in particular to study the stability of the model considered.  Finally, in Section 6, we write the Conclusions of this work.

\section{ DE Model With Higher Time Derivatives Of The Hubble Parameter In Chameleon Brans-Dicke Cosmology}
In this Section, we describe the main cosmological properties of the Chameleon Brans-Dicke (BD) Cosmology, we introduce and describe the DE energy density model we are considering and we introduce the ansatz we have chosen for some of the quantities involved in the following equations.\\
The Chameleon BD model in which the scalar field is coupled non-minimally to the matter field is described by the action $S$ given by the following equation:
\begin{eqnarray}
S= \frac{1}{2}\int d^4x \sqrt{-g}\left[\phi R - \left(\frac{\omega}{\phi}\right)g^{\mu \nu}\nabla_{\mu}\phi \nabla_{\nu}\phi - 2V + 2f\left( \phi\right)L_m\right], \label{1}
\end{eqnarray}
where  $g^{\mu \nu}$ indicates the metric tensor,     $g$ is the determinant of the metric tensor, $R$ indicates the Ricci scalar, $\phi$ indicates the BD scalar field while $f\left( \phi\right)$ and $V (\phi)$ represent two analytic functions of the BD scalar field $\phi$. The matter Lagrangian density, which is indicated with $L_m$, is coupled with the BD scalar field $\phi$ via the analytical  function $f\left( \phi\right)$, which allows a non-minimal coupling between the matter system and the scalar field. In the limiting case corresponding to $f\left( \phi\right)\equiv 1$, we recover the BD action with potential function $\phi$. Varying the action $S$ given in Eq. (\ref{1}) with respect to the metric $g_{\mu \nu}$ and the BD scalar field $\phi$,  we obtain the following two field equations:
\begin{eqnarray}
\phi G_{\mu \nu} &=& T_{\mu \nu}^{\phi} +f\left( \phi\right) T_{\mu \nu}^{m},  \label{2}\\
\left( 2\omega+3  \right)\Box \phi &+& 2\left(2V - V'\phi \right)= T^m f - 2f' \phi_m, \label{3}
\end{eqnarray}
where $ G_{\mu \nu}$ indicates the Einstein tensor,  $\Box$ indicates the d'Alambertian and it is given by  $\Box = \partial^{\mu} \partial_{\mu}$, $T^m = g^{\mu \nu}T^m_{\mu \nu}$  and the prime $'$ indicates a differentiation with respect to the scalar field $\phi$, i.e. $'=\frac{d}{d\phi}$. Moreover, we have that the term $T_{\mu \nu}^{\phi} $ is given by the following relation:
\begin{eqnarray}
T_{\mu \nu}^{\phi} = \frac{\omega}{\phi}\left( \nabla_{\mu}\phi \nabla_{\nu}\phi  - \frac{1}{2}g_{\mu \nu} \nabla_{\alpha}\phi \nabla^{\alpha}\phi  \right) + \left( \nabla_{\alpha} \nabla^{\alpha}\phi  - g_{\mu \nu} \Box \phi\right) -V\left( \phi\right)g_{\mu \nu}  \label{4},
\end{eqnarray}
while the term $T_{\mu \nu}^{m} $ is given by the following relation:
\begin{eqnarray}
T_{\mu \nu}^{m} = \left(\frac{-2}{\sqrt{-g}}\right)\frac{\delta \left( \sqrt{-g}L_m  \right)}{\delta g^{\mu \nu}}. \label{5}
\end{eqnarray}
The explicit coupling between matter system and $\phi$ implies that the stress tensor $T^m_{\mu \nu}$ is not divergence
free.\\
We now apply the above framework to a homogeneous and isotropic Universe described by the Friedman-Robertson-Walker (FRW) metric given by the following relation:
\begin{eqnarray}
ds^2 = -dt^2 + a^2\left( t \right)\left( \frac{dr^2}{1-kr^2} + r^2d\Omega^2   \right),\label{6}
\end{eqnarray}
where $a\left(t\right)$ represents the scale factor (which provides the information about the expansion history of the Universe), $r$ indicates the radial component of the metric, $t$ is the cosmic time and $k$ is the curvature parameter, which can assume the values $-1$, $0$ or $+1$  yielding, respectively, an open, a flat, or a closed FRW Universe. Moreover, the term $d\Omega^2= r^2 \left(d\theta ^2 + \sin^2 \theta d\varphi ^2\right)$ denotes the solid angle element squared. Moreover, we have that the quantities $\theta$ and $\varphi$ are, respectively, the usual azimuthal and polar angles, with $0\leq \theta \leq \pi$ and $0\leq \varphi \leq 2\pi$. \\
In a spatially flat Universe, i.e. for $k=0$, Eqs. (\ref{2}) and (\ref{5}) yield  the following set of three equations:
\begin{eqnarray}
3H^2 &=& \left(\frac{f}{\phi}\right)\rho + \left(\frac{\omega}{2}\right)\frac{\dot{\phi}^2}{\phi}-3H\left(\frac{\dot{\phi}} {\phi}\right), \label{7}\\
3\left(\dot{H} +H^2 \right) &=& -\frac{3\rho}{\phi \left( 2\omega+3  \right)}\left\{ \gamma \phi f' + \left[ \omega \left( \gamma +\frac{1}{3}   \right) +1 \right]f    \right\} \nonumber \\
&&- \left(\frac{\omega}{2}\right)\frac{\dot{\phi}^2}{\phi} +3H\left(\frac{\dot{\phi}} {\phi}\right) + \left(\frac{1}{2\omega +3}\right) \left[3V' + \left( 2\omega -3  \right)\left(\frac{V}{\phi}\right)   \right],\label{8} \\
 \left( 2\omega +3  \right)\left(\ddot{\phi} +3H\dot{\phi}   \right)&-&2\left(2V -\phi V'   \right) = \rho \left[\left(  1-3\gamma \right)f + 2\gamma \phi f'    \right].\label{9}
\end{eqnarray}
We must underline that a prime $'$ indicates a derivative with respect to the scalar field $\phi$ while a dot indicates a derivative with respect to the time $t$, i.e $\dot{}=\frac{d}{dt}$.
Moreover,  in Eqs. (\ref{7}), (\ref{8}) and (\ref{9}), we have that $\rho = \rho_D + \rho_m$ and $\gamma = \frac{p_D}{\rho_m + \rho_D}$. We must also remember that we are considering pressureless DM, then $p_m=0$.\\
In this paper, we investigate the case corresponding to a special ansatz of $V$, $f$, $a$ and $\phi$. Based on this ansatz, we can also determine some constraints on the BD parameter $\omega$.  \\
Following the procedure done in \cite{33},  we choose the following ansatz for the scale factor $a$, the BD scalar field $\phi$ and the analytical functions $V$ and $f$:
\begin{eqnarray}
a\left( t \right) &=& a_0 t^n, \label{12}\\
\phi \left( t \right) &=& \phi_0 t^m, \label{13} \\
V \left( \phi \right) &=& V_0 \phi ^{l_1} = V_0 \left( \phi_0 t^m  \right)^{l_1},  \label{10} \\
f \left( \phi \right) &=& f_0 \phi ^{l_2} = f_0 \left( \phi_0 t^m  \right)^{l_2},
  \label{11}
\end{eqnarray}
where the four quantities $n$, $m$, $l_1$ and $l_2$, are four dimensionless parameters, with $n>0$ in order to have an accelerated Universe. Moreover, $a_0$, $\phi_0$, $V_0$ and $f_0$ indicate, respectively, the present day values of $a$, $\phi$, $V$ and $f$.\\
Since $M_p^2 = \frac{1}{8\pi G}$ and, in BD Chameleon theory, we derive that the BD scalar field $\phi$ is inversely proportional to the Newton's gravitational constant $G$, i.e. we have $\phi \propto G^{-1}$, we can write the energy density of DE $\rho_D$ given in Eq. (\ref{model}) in the framework of Chameleon BD Cosmology as follows:
\begin{eqnarray}
\rho_D = 3\phi \left[ \alpha \left(\frac{\ddot{H}} {H}\right) + \beta H^2 + \varepsilon \dot{H}    \right].\label{rhod}
\end{eqnarray}
We can now derive the expression of some cosmological quantities starting from the expression of the energy density of DE $\rho_D$ given in Eq. (\ref{rhod}).

\subsection{Non Interacting Case}
\begin{figure}[ht]
\centering\includegraphics[width=8cm]{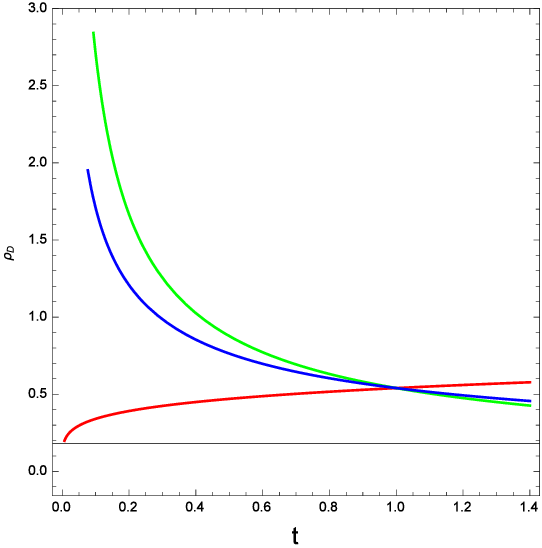}
\caption{Plot of the energy density of DE $\rho_{D}$ given in Eq. (\ref{rhod2}) as function of the cosmic time $t$ for the non interacting case. } \label{rhodnon}
\end{figure}

\begin{figure}[ht]
\centering\includegraphics[width=8cm]{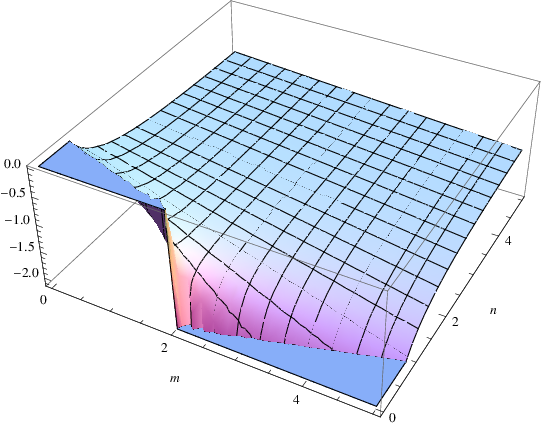}
\caption{Tridimensional plot of the Equation of State (EoS) parameter of DE $w_D$ given in Eq. (\ref{car3}) as function of the parameters  $m$ and $n$ for the non interacting case. We have that $w_D =-1$ for $m=2$ independently on the value of $n$. } \label{omegacham}
\end{figure}

\begin{figure}[ht]
\centering\includegraphics[width=8cm]{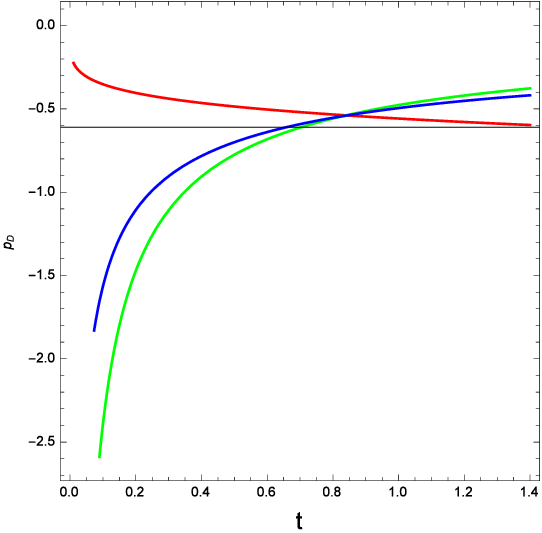}
\caption{Plot of the pressure of DE $p_{D}$ given in Eq. (\ref{15}) as function of the cosmic time $t$ for the non interacting case. } \label{pressurednon}
\end{figure}

We start considering the non interacting case.\\
Using the definition of the scale factor $a\left(t\right)$ given in Eq. (\ref{12}), we obtain the following expression for the Hubble parameter $H$ as function of the cosmic time:
\begin{eqnarray}
H = \frac{\dot{a}} {a} = \frac{n}{t}. \label{hubl}
\end{eqnarray}
In order to obtain a well-defined expression of the Hubble parameter $H$ obtained in Eq. (\ref{hubl}), we must have that $n>0$, which is consistent with the choice previously made to choose positive values of $n$.\\
Differentiating with respect to the cosmic time $t$ the expression of $H$ derived in Eq. (\ref{hubl}), we have that the first and the second time derivatives of the Hubble parameter $H$ are given, respectively, by the following relations:
\begin{eqnarray}
\dot{H} &=&  -\frac{n}{t^2},   \label{dothub}\\
\ddot{H} &=&  \frac{2n}{t^3}. \label{dotdothub}
\end{eqnarray}
Then, using in the definition of the energy density of DE $\rho_D$ given in Eq. (\ref{rhod}) the results for $H$, $\dot{H}$ and $\ddot{H}$ obtained in Eqs. (\ref{hubl}), (\ref{dothub}) and (\ref{dotdothub}) along with the definition of the BD scalar field $\phi$ given in Eq. (\ref{13}),  we can rewrite the expression of the energy density of DE $\rho_D$ as function of the cosmic  time $t$ as follows:
\begin{eqnarray}
\rho_D = 3\phi_0 \left[ 2\alpha + n\left( n\beta - \varepsilon   \right)  \right]t^{m-2}. \label{rhod2}
\end{eqnarray}
We can clearly observe that the expression of $\rho_D$ obtained in Eq. (\ref{rhod2}) depends on the values of the parameters characterizing the various quantities considered. In particular, the behavior of $\rho_D$  depends on the value of $m-2$: if this quantity is positive, then $\rho_D$ has an increasing behavior, if $m-2$ assumes a negative value, then $\rho_D$ has a decreasing behavior.\\
In Figure \ref{rhodnon}, we plot the behavior of the energy density of DE $\rho_D$ for the non interacting case given in Eq. (\ref{rhod2}). We must underline that for this and all the following Figures, we have chosen three different cases: $m=2.2$ (plotted in red), $m=1.5$ (plotted in blue) and $m=1.3$ (plotted in green).  All the other parameters assumes the following values: $a_0=1.1$, $n=2$, $\alpha =2$, $\beta = 1.5$, $\varepsilon = 0.5$ and $\phi_0 = 0.02$  for all the three cases considered. \\
We can clearly observe in Figure \ref{rhodnon} that $\rho_D$ has an increasing behavior for the case corresponding to $m=2.2$ while it has a decreasing behavior for the other two cases considered.\\
We also obtain that, in the limiting case of $m=2$, the energy density of DE $\rho_D$ obtained in Eq. (\ref{rhod2}) assumes a constant value given by the following relation:
\begin{eqnarray}
\rho_D = 3\phi_0 \left[ 2\alpha + n\left( n\beta - \varepsilon   \right)  \right]. \label{rhod2lim1}
\end{eqnarray}
From Eq. (\ref{rhod2lim1}), in order to have a positive defined expression of $\rho_D$, we must have that $2\alpha + n\left( n\beta - \varepsilon   \right) >0$ since  we  assumes that the quantity $\phi_0$ is positive defined. \\
In order to respect the local energy-momentum conservation law, given by the relation $\nabla_{\mu}T^{\mu \nu}=0$ (where the term  $T^{\mu \nu}$ represents the energy-momentum tensor), we have that the total energy density, defined as $\rho_{tot}= \rho_D + \rho_m$, must satisfy the following continuity relation:
\begin{eqnarray}
    \dot{\rho}_{tot}+3H\left( \rho_{tot}+p_{tot} \right)=0,\label{39old}
\end{eqnarray}
where $p_{tot}$ represents the total  pressure.\\
We also have that Eq. (\ref{39old}) can be written as follows:
\begin{eqnarray}
    \dot{\rho}_{tot}+3H\left( 1+w_{tot} \right)\rho_{tot}=0,\label{39}
\end{eqnarray}
where the quantity $w_{tot} \equiv p_{tot}/\rho_{tot}$ represents the total EoS parameter, with $p_{tot}$ being the total pressure, which is equivalent to the DE pressure $p_D$ since we consider pressureless DM.\\
If there is not interaction between DE and DM, the two energy densities $\rho_D$ and $\rho_m$ for DE and DM  are conserved separately according to the following relations:
\begin{eqnarray}
\dot{\rho}_{D}&+&3H\rho_{D}\left(1+w_{D}\right)=0, \label{40} \\
\dot{\rho}_m&+&3H\rho_m= 0,\label{41}
\end{eqnarray}
where $w_{D}$  indicates the EoS parameter of DE. We also have that the EoS parameter of DM $w_m$ is equal to zero since we consider $p_m=0$.\\
From the continuity equation of DE defined in Eq. (\ref{40}), we obtain the following expression for the EoS parameter of DE $w_{D}$:
\begin{eqnarray}
w_{D} &=& -1 - \frac{\dot{\rho}_{D}} {3H\rho_{D}}  . \label{car1}
\end{eqnarray}
We have already obtained the expressions of $H$ and $\rho_D$ respectively in Eqs. (\ref{hubl}) and (\ref{rhod2}), we now need to find the expression of $\dot{\rho}_D$ in order to find the final expression of $w_D$.\\
Using the expression of $\rho_D$ given in Eq. (\ref{rhod2}), we have that the first time derivative of the energy density of DE $\rho_D$ is given by the following relation:
\begin{eqnarray}
\dot{\rho}_D = 3\left(  m-2 \right)\phi_0 \left[ 2\alpha + n\left( n\beta - \varepsilon   \right)  \right]t^{m-3}.\label{car2}
\end{eqnarray}
Therefore, using in Eq. (\ref{car1}) the expressions of $H$, $\rho_D$ and $\dot{\rho}_D$ obtained, respectively, in Eqs. (\ref{hubl}), (\ref{rhod2}) and (\ref{car2}),  we obtain that the final expression of the EoS parameter of DE $w_D$  is given by the following relation:
\begin{eqnarray}
w_{D} &=& -\frac{3n+m-2}{3n} \nonumber \\
&=& -1 -\frac{m-2}{3n} , \label{car3}
\end{eqnarray}
i.e. $w_D$ assumes a constant value which depends on the values of the exponents $n$ and $m$. In particular, we obtain that $w_D = -1$ for $m=2$ independently on the value of $n$.\\
In Figure \ref{omegacham},   we plot the behavior of the EoS parameter of DE $w_D$ obtained in Eq. (\ref{car3}) for a range of values of the exponents $m$ and $n$. In particular, we have chosen $n>0$ (which is required in order to have an accelerated Universe) and $m\geq 0$. \\
We obtain that the EoS parameter of DE $w_D$ can assume values which can be greater, equals or lower than $-1$ according to the values assumed by $m$ and $n$. In particular, we have that $w_D > -1$ when $m>2$ while we have that $w_D < -1$ when $m<2$.  \\
Using the continuity equation for DE given in Eq. (\ref{40}), we obtain the following general expression for the pressure of DE $p_D$ as follows:
\begin{eqnarray}
p_D = -\rho_D -\frac{\dot{\rho}_D}{3H}.\label{14non}
\end{eqnarray}
Using in Eq. (\ref{14non}) the expression of $\rho_D$ obtained in Eq. (\ref{rhod2}) along with the expressions of $H$ and   $\dot{\rho}_D$ obtained, respectively, in Eqs. (\ref{hubl}) and  (\ref{car2}), we obtain the following expression for $p_D$:
\begin{eqnarray}
p_D =  -\phi_0 \left(\frac{3n+m-2}{n}\right)\left[  2\alpha + n\left( n\beta - \varepsilon   \right)   \right]t^{m-2}.\label{15}
\end{eqnarray}
We clearly observe that, in the limiting case corresponding to $m=2$, the expression of $p_D$ defined in Eq. (\ref{15}) assumes the constant values given by the relation:
\begin{eqnarray}
p_D =  -3\phi_0 \left[  2\alpha + n\left( n\beta - \varepsilon   \right)   \right].\label{15bis}
\end{eqnarray}
Therefore, we obtain that, for $m=2$, we have $p_D = -\rho_D$.\\
In Figure \ref{pressurednon}, we plot the behavior of the pressure of DE $p_D$ for the interacting case given in Eq. (\ref{15}).
Both expressions of $p_D$ obtained in Eqs. (\ref{15}) and  (\ref{15bis}) depend on the values of the parameters characterizing the quantities involved in this paper; moreover,  Eqs. (\ref{15}) and  (\ref{15bis}) can assume both positive and negative values depending on the choice made for the parameters involved. We can clearly observe that for the case corresponding to $m=2.2$, the pressure of DE $p_D$ has a decreasing value while for the other two cases it has an increasing behavior. Moreover, we observe that $p_D$ always assumes negative values. \\
We can also study the effective EoS parameter $w_{eff}$, which is generally defined as follows:
\begin{eqnarray}
w_{eff} = \frac{p_D}{\rho_D+\rho_m}. \label{omegaeffold}
\end{eqnarray}
We must underline also that $\gamma = w_{eff}$.\\
We have already obtained the expressions of the energy density of DE $\rho_D$ and of the pressure of DE $p_D$ respectively, in Eqs. (\ref{rhod2}) and (\ref{15}), we now must derive the expression of the energy density of DM $\rho_m$.\\
From the continuity equation for DM given in Eq. (\ref{41}), we obtain the following expression for $\rho_m$:
\begin{eqnarray}
\rho_m &=& \rho_{m0} a^{-3} \nonumber \\
&=& \rho_{m0}a_0^{-3}t^{-3n},  \label{rhom}
\end{eqnarray}
where $\rho_{m0}$ indicates the present day value of $\rho_m$ and we have used the definition of the scale factor $a$ given in Eq. (\ref{12}). In all Figures where $\rho_{m0} $ is involved, we  consider $\rho_{m0} = 0.3$.   Therefore, using in Eq. (\ref{omegaeffold}) the expressions of $\rho_D$, $p_D$ and $\rho_m$ given, respectively, in Eqs. (\ref{rhod2}), (\ref{15}) and (\ref{rhom}), we obtain the following expression for $w_{eff}$:
\begin{eqnarray}
w_{eff}&=&-\frac{(m+3 n-2) t^{-2+m} \left[2 \alpha +n (n \beta -\epsilon)\right] \phi _0}{n \left\{\frac{t^{-3 n} \rho_{m0}}{a_0^3}+3 t^{-2+m} \left[2\alpha +n \left(n \beta -\epsilon \right)\right] \phi _0 \right\}} \nonumber \\
&=& -\frac{(m+3 n-2)  \left[2 \alpha +n \left(n \beta -\epsilon\right)\right] \phi _0}{n \left\{\frac{t^{-3 n- \left(m-2\right)} \rho_{m0}}{a_0^3}+3  \left[2\alpha +n \left(n \beta -\epsilon \right)\right] \phi_0\right\}} .\label{omegaeff}
\end{eqnarray}
For $w_{eff}$ and all the subsequent quantities we plot, we calculate the values they assume for three different limiting cases: 1) $t\rightarrow 0$, i.e. at the beginning of the Universe 2) $t\rightarrow \infty$, i.e. for far future, 3) at present time, indicated with $t_0$.
We must do some considerations about the value assumed by $t_0$. We now that the scale factor $a\left(t\right)$ is related to the redshift $z$ thanks to the relation:
\begin{eqnarray}
a\left( t \right) = \frac{1}{1+z}.
\end{eqnarray}
Therefore, using the expression of scale factor we have chosen in Eq. (\ref{12}), we can write:
\begin{eqnarray}
t = \left[\left(\frac{1}{1+z}\right)\left(\frac{1}{a_0}   \right)\right]^{\frac{1}{n}}.
\end{eqnarray}
The present time $t_0$ is obtained for $z=0$, therefore we have that $t_0$ can be obtained using the relation:
\begin{eqnarray}
t_0= \left(\frac{1}{a_0}\right)^{\frac{1}{n}}.
\end{eqnarray}
Considering the values of $a_0$ and $n$ we are considering in this paper, we obtain that:
\begin{eqnarray}
t_0\approx 0.953.
\end{eqnarray}
For $t\rightarrow 0$, we have that $\omega_{eff}\rightarrow 0$ for all the cases considered in this paper.\\
In the limiting case of $t\rightarrow \infty$, i.e. for far future, we obtain from Eq. (\ref{omegaeff}) that $w_{eff} = w_D$. Considering the expression of $w_D$ obtained in Eq. (\ref{car3}), we obtain for the three different cases we are considering that:
\begin{eqnarray}
w_{eff,late,1} &\approx& -1.033,\\
w_{eff,late,2} &\approx& -0.917,\\
w_{eff,late,3} &\approx&-0.883.
\end{eqnarray}
At present time, i.e. for $t=t_0$, we obtain, for the three different cases we are considering, that:
\begin{eqnarray}
w_{eff,present,1} &\approx& -0.662,\\
w_{eff,present,2} &\approx& -0.594,\\
w_{eff,present,3} &\approx&-0.575.
\end{eqnarray}

\begin{figure}[ht]
\centering\includegraphics[width=8cm]{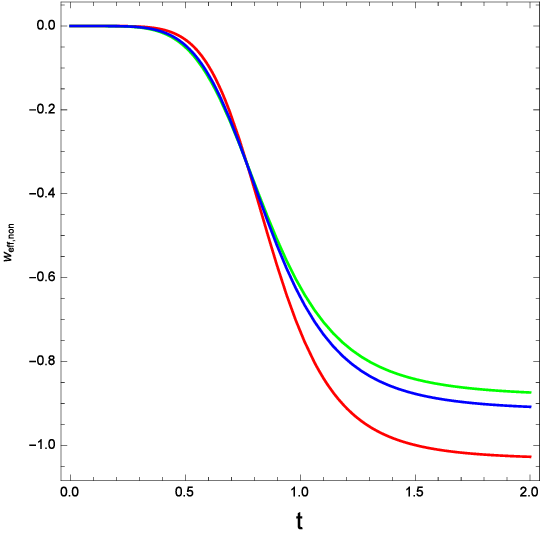}
\caption{Plot of the effective Equation of State (EoS) parameter $w_{eff}$ given in Eq. (\ref{omegaeff}) as function of the cosmic time $t$ for the non interacting case. } \label{omegaeffnon}
\end{figure}

\begin{figure}[ht]
\centering\includegraphics[width=8cm]{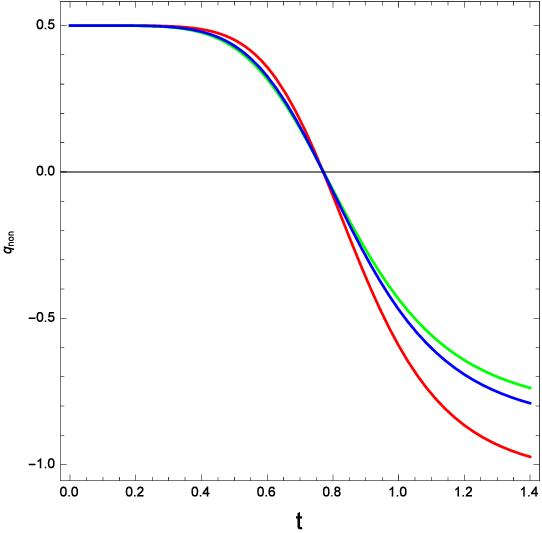}
\caption{Plot of the deceleration parameter $q_{non}$ given in Eq. (\ref{decelnon1}) as function of the cosmic time $t$ for the non interacting case. } \label{qnon}
\end{figure}

In Figure \ref{omegaeffnon},  we plot the behavior of $w_{eff}$ given in Eq. (\ref{omegaeff}).
We can clearly observe that, initially, $w_{eff}$ has a monotone decreasing behavior for all the three cases considered, while it starts to have a flat behavior for later times tending to $w_{eff}\rightarrow \approx -1$. We also obtain, as expected, that for $t\rightarrow 0$ $\omega_{eff}\rightarrow 0$.  Moreover, for all the cases considered, we start with $w_{eff}>-1$, but it has a different behavior for later times according to the value of $m$ considered: for the case with $m=2.2$, we have that for $t\approx 1.4$ $w_{eff}$ crosses the phantom boundary, instead for the other two cases we have that $w_{eff}$ asymptotically tends to -1 without crossing it.\\
Therefore,  for the non interacting case, we conclude that $w_{eff}$ has a quintom-like behavior for $m=2.2$.\\
In the limiting case of $m=2$, we obtain that:
\begin{eqnarray}
w_{eff}&=& -\frac{  3\left[2 \alpha +n \left(n \beta -\epsilon\right)\right] \phi _0}{ \frac{t^{-3 n} \rho_{m0}}{a_0^3}+3  \left[2\alpha +n \left(n \beta -\epsilon \right)\right] \phi_0} .\label{omegaefflim2}
\end{eqnarray}
For $t\rightarrow \infty$, we derive from Eq. (\ref{omegaefflim2}) that $w_{eff}\rightarrow -1$.\\
Furthermore, for all the set of values given by $m=2-3n$, we obtain that $w_{eff}=0$ independently on the values of the other parameters involved. \\
We now want to derive some information about the BD parameter $\omega$.\\
Using the results of Eqs. (\ref{rhod2}), (\ref{15}) and (\ref{rhom}) in Eq. (\ref{8}), we obtain the following relation:
\begin{eqnarray}
-\left(\frac{9f_0\phi_0^{l_2}} {2\omega + 3}\right)\left[ 2\alpha + n\left( n\beta - \varepsilon   \right)  \right]\left[ \gamma l_2 + \omega \left(\gamma + \frac{1}{3}   \right)+1  \right]t^{ml_2} \nonumber \\
-\left\{\frac{3\rho_{m0}f_0\phi_0^{l_2-1}\left[ \gamma l_2 + \omega \left(\gamma + \frac{1}{3}
   \right)+1    \right]}{a_0^3\left(2\omega +3   \right)}\right\}t^{m\left(l_2 -1 \right) + 2-3n} \nonumber \\
+\left[\frac{V_0\left(3l_1 + 2\omega -3  \right)\phi_0^{l_1-1}} {2\omega+3}\right]t^{2+m\left( l_1 -1 \right)} = 3n\left( n-1\right) -\omega n^2 + 2nm \label{ref},
\end{eqnarray}
where we used the following relations:
\begin{eqnarray}
\dot{\phi} &=& \frac{d\phi}{dt} = m\phi_0 t^{m-1},  \label{}\\
V' &=& \frac{dV}{d\phi} = V_0 l_1 \phi^{l_1-1} = V_0 l_1 \phi_0^{l_1-1}t^{m\left(l_1 -1  \right)},  \label{}\\
f' &=& \frac{df}{d\phi} = f_0 l_2 \phi^{l_2-1} = f_0 l_2 \phi_0^{l_2-1}t^{m\left(l_2 -1  \right)}.\label{}
\end{eqnarray}
We must also underline that, using the expression of $\gamma$ equivalent to $w_{eff}$, there are no further temporal contribution to the terms of Eq. (\ref{ref}) where $\gamma$ is present.\\
In Eq. (\ref{ref}), we have that the right hand side term, i.e. $3n\left( n-1\right) -\omega n^2 + 2nm$, is a constant term while all the terms in the left side have a temporal dependance. Therefore, we have that the right and the left sides can be equals only if all the exponents  of $t$ are equals to zero. We have that the exponent of the second and third terms of the left hand side can be equals to zero, while we have that $ml_2 \neq 0$ for all values of $m$ and $l_2$ we are considering. Then, we have that its coefficient must be equal to zero. We have two possibilities which can satisfy this condition:
\begin{eqnarray}
2 \alpha +n (n \beta -\epsilon) &=&0, \label{kir1}\\
\gamma l_2 + \omega \left(\gamma + \frac{1}{3}   \right)+1 &=&0. \label{kir2}
\end{eqnarray}
The condition given in Eq. (\ref{kir1}) leads to $p_D=0$, therefore we exclude it and we consider only the condition given in Eq. (\ref{kir2}).\\
Therefore, from Eq. (\ref{ref}), we obtain the following restrictions on the parameters involved:
\begin{eqnarray}
\omega &=& -\frac{3\left( 1+\gamma l_2  \right)}{1+3 \gamma},  \label{rituccia1}\\
-2  &=&   m\left( l_2 -1  \right) -3n,  \label{rituccia2}\\
-2  &=&  m\left( l_1 -1  \right).   \label{rituccia3}
\end{eqnarray}
Moreover, combining the results of Eqs. (\ref{rituccia2}) and (\ref{rituccia3}), we obtain the following condition:
\begin{eqnarray}
m\left( l_2 - l_1  \right)   &=&  3n.  \label{rituccia3lu}
\end{eqnarray}
Furthermore, combining Eqs. (\ref{rituccia1}) and (\ref{rituccia3lu}), we obtain the following relation for $\omega$:
\begin{eqnarray}
\omega &=& -\frac{3\left[ 1+\gamma\left( \frac{3n}{m}+ l_1  \right)   \right]}{1+3 \gamma}.  \label{rituccia1emilio}
\end{eqnarray}
From Eq. (\ref{rituccia3}), we obtain that $l_1$ is given by:
\begin{eqnarray}
l_1 = \frac{2}{m}+1.
\end{eqnarray}
Therefore, we can write the BD parameter $\omega$ as follows:
\begin{eqnarray}
\omega &=& -\frac{3\left[ 1+\gamma\left( \frac{3n-2}{m}+1  \right)   \right]}{1+3 \gamma}.  \label{rituccia1emilio2}
\end{eqnarray}
In 1973, the condition $\omega > 5$ was consistent with data known at the epoch. By 1981, the constrain $\omega > 30$ was consistent with  data available that time. In 2003, evidence derived from the Cassini-Huygens experiment indicates that the value of $\omega$ must exceed 40,000 \cite{53}. The information and constrains on the value of $\omega$ can also help to have better constrains on the values of the other parameters involved thanks to the result of Eq. (\ref{rituccia1}).  \\
We can also study the behavior of the deceleration parameter $q$, which can be obtained from the general relation given by:
\begin{eqnarray}
q &=&  -\frac{\ddot{a}a}{\dot{a}^2} -1 - \frac{\dot{H}}{H^2}. \label{43}
\end{eqnarray}
The expansion of the Universe results to be accelerated if the term $\ddot{a}$ has a positive value, as recent cosmological measurements suggest;  in this case, $q$ assumes a negative value.\\
We have that the deceleration parameter $q$ can be also written as function of the total pressure $p_{tot}$ and the total energy density $\rho_{tot}$ as follows:
\begin{eqnarray}
q&=& \frac{1}{2} + \frac{3}{2}\left(\frac{p_{tot}}{\rho_{tot}}\right) \nonumber \\
&=& \frac{1}{2} + \frac{3}{2}\left(\frac{p_D}{\rho_D +\rho_m}\right).  \label{decelnon}
\end{eqnarray}
Using in Eq. (\ref{decelnon}) the expression of $\rho_D$, $p_D$ and $\rho_m$ obtained in Eqs. (\ref{rhod2}), (\ref{15}) and (\ref{rhom}), we obtain the following final expression for the deceleration parameter for the non interacting case $q_{non} $:
\begin{eqnarray}
q_{non}  &=&  \frac{1}{2} - \frac{3}{2}\left\{\frac{ \phi_0 \left(3n+m-2\right)\left[  2\alpha + n\left( n\beta - \varepsilon   \right)   \right]t^{m-2}} {3n\phi_0 \left[ 2\alpha + n\left( n\beta - \varepsilon   \right)  \right]t^{m-2} + \rho_{m0}na_0^{-3}t^{-3n} }\right\} \nonumber \\
&=&    \frac{1}{2} - \frac{3}{2}\left\{\frac{ \phi_0 \left(3n+m-2\right)\left[  2\alpha + n\left( n\beta - \varepsilon   \right)   \right]}{3n\phi_0 \left[ 2\alpha + n\left( n\beta - \varepsilon   \right)  \right] + \rho_{m0}na_0^{-3}t^{-\left(3n +m -2 \right)} }\right\} . \label{decelnon1}
\end{eqnarray}
In Figure \ref{qnon},  we plot the behavior of $q_{non}$ obtained in Eq. (\ref{decelnon1}).
We can observe that $q_{non}$ has a decreasing behavior for all the three cases considered, starting with $q_{non}>0$. The transition between non accelerated and accelerated Universe, i.e. between $q_{non}>0$ and $q_{non}<0$, happens at $t \approx 0.7 \div 0.8$, depending on the values of the parameters involved. We observe that for $t\rightarrow 0$, we obtain  $q_{non} \approx 1/2$ for all the cases considered (as it was already derived in Figure \ref{qnon}), which means that at the the beginning of the cosmic history the models considered lead to a decelerated Universe. \\
Instead, in the limiting case of $t\rightarrow \infty$,  we obtain that $q_{non}$ is given by the following expression:
\begin{eqnarray}
q_{non,late}  &=&   \frac{1}{2}\left[ 1- \frac{  \left(3n+m-2\right)}{n  }   \right] \nonumber \\
&=&-1- \frac{  \left(m-2\right)}{2n  }  \label{decelnon1lim},
\end{eqnarray}
which is a constant value depending on the values of the exponents $n$ and $m$ only.\\
Considering the values of the parameters we considered for the plots,   we obtain the following values for the three different cases we take into account:
\begin{eqnarray}
q_{non,late,1}   &\approx&-1.050  \label{decelnon1lim1},\\
q_{non,late,2}   &\approx&-0.875  \label{decelnon1lim2},\\
q_{non,late,3}   &\approx&-0.825  \label{decelnon1lim3}.
\end{eqnarray}
Therefore, we obtain values of the deceleration parameter which indicate an accelerated Universe for the far future.\\
At present time, we obtain that:
\begin{eqnarray}
q_{non,present,1}   &\approx&  -0.493, \label{decelnon1lim1pr}\\
q_{non,present,2}   &\approx&  -0.391, \label{decelnon1lim2pr}\\
q_{non,present,3}   &\approx& -0.362,  \label{decelnon1lim3pr}
\end{eqnarray}
therefore at present time we obtain values of the deceleration parameter which indicate an accelerated Universe, result in agreement with the available cosmological observation.\\
In the limiting case of $m+3n-2=0$, or equivalently $m = 2-3n$, we obtain from Eq. (\ref{decelnon1lim}) that $q=1/2$, which means a decelerated Universe, result which is in contrast with the present day observations. \\
The limiting case corresponding to $q_{non}=0$, which indicates the transition between decelerated and accelerated Universe, happens for Eq. (\ref{decelnon1}) when:
\begin{eqnarray}
\left\{\frac{ \phi_0 \left(3n+m-2\right)\left[  2\alpha + n\left( n\beta - \varepsilon   \right)   \right]}{3n\phi_0 \left[ 2\alpha + n\left( n\beta - \varepsilon   \right)  \right] + \rho_{m0}a_0^{-3}t^{-\left(3n+m-2\right)} }\right\} = \frac{1}{3}. \label{decelnon1lim2}
\end{eqnarray}
Therefore, we will have accelerated Universe (i.e. $q<0$) from Eq. (\ref{decelnon1}) if:
\begin{eqnarray}
\left\{\frac{ \phi_0 \left(3n+m-2\right)\left[  2\alpha + n\left( n\beta - \varepsilon   \right)   \right]}{3n\phi_0 \left[ 2\alpha + n\left( n\beta - \varepsilon   \right)  \right] + \rho_{m0}a_0^{-3}t^{-\left(3n+m-2\right)} }\right\} < \frac{1}{3}. \label{decelnon1lim3}
\end{eqnarray}
For $t\rightarrow 0$, we have an accelerated Universe provided that:
\begin{eqnarray}
-1- \frac{  \left(m-2\right)}{2n  } <0,
\end{eqnarray}
which implies that:
\begin{eqnarray}
m>2\left( n-1\right).
\end{eqnarray}

\subsection{Interacting Case}
\begin{figure}[ht]
\centering\includegraphics[width=8cm]{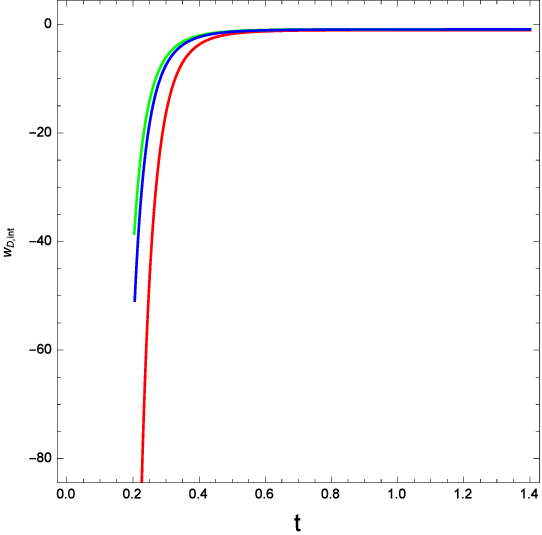}
\caption{Plot of the Equation of State (EoS) parameter of DE $w_{D,int}$ given in Eq. (\ref{car3int2}) as function of the cosmic time $t$ for the interacting case. } \label{omegadint}
\end{figure}

\begin{figure}[ht]
\centering\includegraphics[width=8cm]{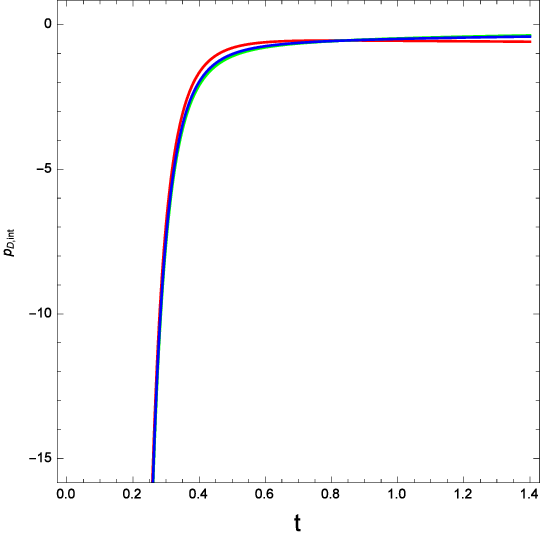}
\caption{Plot of the pressure of DE $p_{D,int}$ given in Eq. (\ref{14int1}) as function of the cosmic time $t$ for the interacting case. } \label{pressuredint}
\end{figure}

\begin{figure}[ht]
\centering\includegraphics[width=8cm]{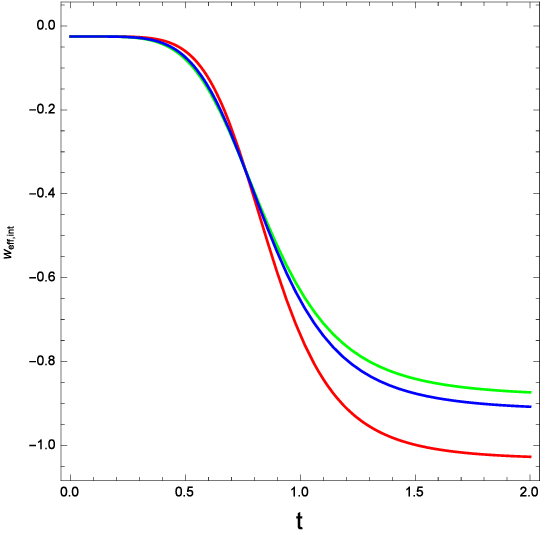}
\caption{Plot of the effective Equation of State (EoS) parameter $w_{eff,int}$ obtained in Eq. (\ref{gollumeff}) as function of the cosmic time $t$ for the interacting case. } \label{omegaeffint}
\end{figure}

We now extend the calculations accomplished in the previous Subsection to the case of presence of interaction between the two Dark Sectors. \\
Some recent observational evidences obtained about the cluster of galaxies known as Abell A586 clearly support the existence of a kind of interaction between DE and DM \cite{q2,q2-2}.  Unfortunately, the precise strength of this interaction is not clearly determined \cite{q3}. The existence of an interaction between the Dark Sectors can be also detected during the formation of the Large Scale Structures (LSS).  It was considered that the dynamical equilibrium of collapsed structures (like clusters of galaxies) can be modified because of the coupling and interaction between DE and DM \cite{1abd,10abd}. The main idea is that the virial theorem results to have a modification due to the energy exchange between DM and DE, which leads to a bias in the estimation of the virial masses of clusters of galaxies when the usual virial conditions are considered. Other observational signatures on the Dark Sectors mutual interaction can be observed in the probes of the cosmic expansion history by using results of the Baryonic Acoustic Oscillation (BAO), Supernovae Ia (SNeIa)  and CMB shift data \cite{A173,22abd}.\\
In presence of interaction between the two Dark Sectors, the energy densities of DE and DM $\rho_D$ and $\rho_m$ are conserved separately and the conservation equations take the following form:
\begin{eqnarray}
\dot{\rho}_{D}&+&3H\rho_{D}\left(1+w_{D}\right)=-Q, \label{46} \\
\dot{\rho}_m&+&3H\rho_m=Q .\label{47}
\end{eqnarray}
We have that in Eqs. (\ref{46}) and (\ref{47}), the term $Q$ represents an interaction term which is an arbitrary function of cosmological parameters, like the Hubble parameter $H$, the deceleration parameter $q$ and the energy densities of DM and DE $\rho_m$ and $\rho_D$, i.e. $Q(\rho_m,\rho_D,H,q)$.
Many different candidates have been proposed in order to describe $Q$. In this paper, we have chosen to consider the following one:
\begin{eqnarray}
Q_1 &=& 3b^2H\rho_m  ,\label{48}
\end{eqnarray}
where the term $b^2$ represents a coupling parameter between DM and DE, which is also known as transfer strength or interaction parameter \cite{q1,q1-4,q1-8}.
Thanks to the observational cosmological data obtained from the Gold SNe Ia samples, the CMB data from the WMAP satellite and the Baryonic Acoustic Oscillations (BAO) from the Sloan Digital Sky Survey (SDSS), it is established that  the coupling parameter between DM and DE must assume a small positive value, which is in agreement with the requirements for solving the cosmic coincidence problem and the constraints which are given by the second law of thermodynamics \cite{feng08}.
Observations of CMB radiation and of clusters of galaxies  suggest that $0<b^2 < 0.025$  \cite{q4}. This result  is in agreement with the fact that the interaction term $b^2$ must be taken in the range of values [0,1] \cite{zhang-02-2006}  with the limiting case of $b^2 = 0$ leading to the non-interacting FRW model. We must also remember and underline that other interaction terms have been proposed and well studied in literature.  \\
From the continuity equation for DE defined in Eq. (\ref{40}), we obtain the following general  expression for the EoS parameter of DE $w_{D}$:
\begin{eqnarray}
w_{D} &=& -1 - \frac{\dot{\rho}_D}{3H\rho_{D}} - \frac{Q}{3H\rho_D}  . \label{car1int}
\end{eqnarray}
We have already obtained the expressions of $\rho_D$ and $\dot{\rho}_{D}$, respectively, in Eqs. (\ref{rhod2}) and (\ref{car2}), we now need to find the expression of $\rho_m$ for the interacting case in order to be able to obtain the final expression of the EoS parameter of DE.\\
Solving the continuity equation for DM given in Eq. (\ref{47}), we obtain the following expression for $\rho_{m,int}$:
\begin{eqnarray}
\rho_{m,int} &=& \rho_{m0} a^{-3\left( 1-b^2  \right)} \nonumber \\
&=& \rho_{m0}a_0^{-3\left( 1-b^2  \right)}t^{-3\left( 1-b^2  \right)n}, \label{rhomint}
\end{eqnarray}
where we have used the expression of the scale factor $a(t)$ defined in Eq. (\ref{12}).  We have that, in the limiting case corresponding to $b^2=0$ (i.e. in absence of interaction), we recover the same  result of the  non interacting case obtained in the previous subsection.\\
Therefore, using the expressions of the Hubble parameter $H$, the energy density of DE  $\rho_D$ and the energy density of DM$\rho_{m,int}$ obtained, respectively, in Eqs.  (\ref{hubl}), (\ref{rhod2}) and (\ref{rhomint}), we derive the following expression for the EoS parameter for the interacting case $w_{D,int} $:
\begin{eqnarray}
w_{D,int} &=& -\frac{3n+m-2}{3n}  - \frac{b^2 \rho_{m0}a_0^{-3\left( 1-b^2  \right)}t^{-3\left( 1-b^2  \right)n-\left( m-2 \right)}} {3\phi_0 \left[ 2\alpha + n\left( n\beta - \varepsilon   \right)  \right]},  \label{car3int}
\end{eqnarray}
which can be also written as follows:
\begin{eqnarray}
w_{D,int} &=& w_{D}  - \frac{b^2 \rho_{m0}a_0^{-3\left( 1-b^2  \right)}t^{-3\left( 1-b^2  \right)n-\left( m-2\right)}} {3\phi_0 \left[ 2\alpha + n\left( n\beta - \varepsilon   \right)  \right]}. \label{car3int2}
\end{eqnarray}
We have that, in the limiting case corresponding to $b^2=0$ (i.e. in absence of interaction), we recover the result of the  non interacting case, i.e. $w_{D,int} = w_{D}$.\\
In Figure \ref{omegadint}, we plot the behavior of the EoS parameter of DE for the interacting case $w_{D,int}$ given in Eq. (\ref{car3int2}).
We observe  that $w_{D,int}$ has an increasing behavior for all the cases considered, starting from values much lower than -1. For $t\rightarrow 0$, we have that $w_{D,int}\rightarrow -\infty$. Instead, for $t\rightarrow \infty$, we obtain that $w_{D,int} \rightarrow w_{D}$. Therefore, considering the expression of $w_{D}$ given in Eq. (\ref{car3}), we obtain the following set of values of the three cases considered:
\begin{eqnarray}
w_{D,int,late,1} &\approx& -1.033,\\
w_{D,int,late,2} &\approx& -0.917,\\
w_{D,int,late,3} &\approx& -0.883.
\end{eqnarray}
At present time, we obtain the following values of the EoS parameter of DE $w_{D,int}$ for the interacting case for the three cases considered:
\begin{eqnarray}
w_{D,int,present,1}   &\approx& -1.047, \label{}\\
w_{D,int,present,2}   &\approx& -0.930, \label{}\\
w_{D,int,present,3}   &\approx& -0.897.  \label{}
\end{eqnarray}
Therefore, for the case with $m=2.2$, we obtain a value which is beyond the phantom divide line, while for the other two cases we obtain values higher than -1.\\
Moreover, for the first two cases considered, i.e. for $m=2.2$ and $m=1.5$, we have that the value of $w_{D,int}$ we obtained lie within the constraints obtained through observations given in Table 1.\\
Using the conservation equation for DE given in Eq. (\ref{46}), we also obtain the following expression for the pressure of DE $p_D$ for the interacting case:
\begin{eqnarray}
p_{D,int} = -\rho_D -\frac{\dot{\rho}_D}{3H} - \frac{Q}{3H} .\label{14int}
\end{eqnarray}
Using the expressions of $\rho_D$ and $\rho_{m,int}$ derived, respectively, in Eqs. (\ref{rhod}) and (\ref{rhomint}), we find the following solution for $p_D$:
\begin{eqnarray}
p_{D,int} = -\phi_0 \left(\frac{3n+m-2}{n}\right)\left[  2\alpha + n\left( n\beta - \varepsilon   \right)   \right]t^{m-2} -\frac{b^2 t^{-3 \left(1-b^2\right) n} \rho _{m0 }}{a_0^3} \label{14int1},
\end{eqnarray}
which can be also written as follows:
\begin{eqnarray}
p_{D,int} = p_{D,non} -\frac{b^2 t^{-3 \left(1-b^2\right) n} \rho _{m0 }}{a_0^3} .\label{14-2}
\end{eqnarray}
In Figure \ref{pressuredint}, we plot the behavior of the pressure of DE $p_{D,int}$ for the interacting case given in Eq. (\ref{14int1}).
We can observe that $p_D$ has an increasing behavior for all the cases considered, staying always negative.\\
We have that, in the limiting case corresponding to $b^2=0$ (i.e. in absence of interaction), we recover the result of the non interacting case.\\
We can also study the effective EoS parameter $w_{eff}$, which is defined for the interacting case as follows:
\begin{eqnarray}
w_{eff,int} = \frac{p_{D,int}} {\rho_D+\rho_{m,int}} . \label{gollum}
\end{eqnarray}
We have already obtained the expressions of $\rho_D$, $p_{D,int}$ and $\rho_{m,int}$, respectively, in Eqs. (\ref{rhod2}), (\ref{14-2}) and (\ref{rhomint}).  Therefore, the final expression of $w_{eff}$ is given by:
\begin{eqnarray}
w_{eff,int} &=& A_{eff} + B_{eff}, \label{gollumeff}
\end{eqnarray}
where the quantities $A_{eff}$ and $B_{eff}$ are defined as follows:
\begin{eqnarray}
A_{eff} &=& -\frac{b^2 t^{-3 \left(1-b^2\right) n} \rho_{m0}} {t^{-3 \left(1-b^2\right) n} \rho _{m0}+3a_0^3 t^{-2+m} \left[2 \alpha +n \left(n \beta -\epsilon \right)\right] \phi _0}\nonumber \\
&=&-\frac{b^2 t^{-3 \left(1-b^2\right) n-\left(  m-2 \right)} \rho_{m0}} {t^{-3 \left(1-b^2\right) n-\left(  m-2 \right)} \rho_{m0}+3a_0^2 \left[2 \alpha +n \left(n \beta -\epsilon \right)\right] \phi _0}, \\
B_{eff} &=& -\frac{\frac{(-2+m+3 n) t^{m-2} \left[2 \alpha +n (n \beta -\epsilon )\right] \phi_0}{n}} {\frac{t^{-3 \left(1-b^2\right) n} \rho_{m0}} {a_0^3}+3 t^{m-2} \left[2 \alpha +n (n \beta -\epsilon )\right] \phi _0} \nonumber \\
&=&-\frac{\frac{(-2+m+3 n)  \left[2 \alpha +n (n \beta -\epsilon )\right] \phi_0}{n}} {\frac{t^{-3 \left(1-b^2\right) n-\left( m-2\right)} \rho_{m0}} {a_0^3}+3  \left[2 \alpha +n (n \beta -\epsilon )\right] \phi _0}.
\end{eqnarray}
We have that, in the limiting case corresponding to $b^2=0$ (i.e. in absence of interaction), we recover the result of the non interacting case.\\
In Figure \ref{omegaeffint},  we plot the behavior of $w_{eff,int}$ given in Eq. (\ref{gollumeff}).

\begin{figure}[ht]
\centering\includegraphics[width=8cm]{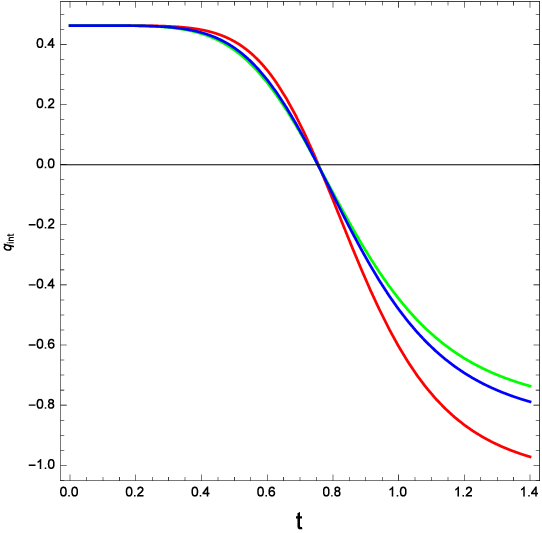}
\caption{Plot of the deceleration parameter $q_{int}$ given in Eq. (\ref{qint!!}) as function of the cosmic time $t$ for the interacting case. } \label{qint}
\end{figure}

We can clearly observe in Figure \ref{omegaeffint} that, initially, $w_{eff,int}$ has a decreasing behavior for all the cases considered, while it starts to have a flatter  behavior for later times. Moreover, for all the cases considered, we start with $w_{eff,int}>-1$, but we can observe that asymptotically tend to the value $-1$. \\
For $t\rightarrow 0$, we obtain that $w_{eff,int} \rightarrow -b^2 = -0.025$, Instead, we have that in the limiting case of $t\rightarrow \infty$, we obtain that $w_{eff,int} = w_{D}$. Therefore, we obtain that:
\begin{eqnarray}
w_{eff,int,late,1} &\approx& -1.033,\\
w_{eff,int,late,2} &\approx& -0.917,\\
w_{eff,int,late,3} &\approx&-0.883.
\end{eqnarray}
At present time, we obtain the following values for the three different cases we consider:
\begin{eqnarray}
w_{eff,int,present,1} &\approx& -0.673,\\
w_{eff,int,present,2} &\approx& -0.605,\\
w_{eff,int,present,3} &\approx&-0.585.
\end{eqnarray}
We conclude, then, that the presence of interaction clearly affect the results we obtain. In fact, the value of  $w_{eff,int}>-1$ for $t\rightarrow 0$ is different from the value obtained for the non interacting case, in particular we obtain a lower value. \\
Finally, using the expression of  $\rho_D$, $\rho_{m,int}$ and $p_{D,int}$ obtained, respectively, in Eqs. (\ref{rhod2}), (\ref{rhomint}) and  (\ref{14-2}), we obtain the following expression for $q_{int} $:
\begin{eqnarray}
q_{int} &=&  \frac{1}{2} - \frac{3}{2}\left\{\frac{ \phi_0 \left(3n+m-2\right)\left[  2\alpha + n\left( n\beta - \varepsilon   \right)   \right]t^{m-2}} {3n\phi_0 \left[ 2\alpha + n\left( n\beta - \varepsilon   \right)  \right]t^{m-2} + n\rho_{m0}a_0^{-3\left( 1-b^2  \right)}t^{-3\left( 1-b^2  \right)n} } \right\} \nonumber \\
&=&  \frac{1}{2} - \frac{3}{2}\left\{\frac{ \phi_0 \left(3n+m-2\right)\left[  2\alpha + n\left( n\beta - \varepsilon   \right)   \right]}{3n\phi_0 \left[ 2\alpha + n\left( n\beta - \varepsilon   \right)  \right] + n\rho_{m0}a_0^{-3\left( 1-b^2  \right)}t^{-\left[3\left( 1-b^2  \right)n +m-2\right]} } \right\}.\label{qint!!}
\end{eqnarray}
In Figure \ref{qint},  we plot the behavior of $q_{int}$ given in Eq. (\ref{qint!!}).\\
We can observe that $q_{int}$ has a decreasing behavior for all the three cases considered, starting with $q_{int}>0$. The transition between non accelerated and accelerated Universe, i.e. between $q_{int}>0$ and $q_{int}<0$, happens at $t \approx 0.7-0.8$, depending on the values of the parameters involved. We can also observe that asymptotically we have $q_{int}\rightarrow -1$ for all the cases considered.\\
We can also make some considerations about limiting values of $q_{int}$. \\
For $t\rightarrow 0$, we obtain the following value of $q_{int,0}$:
\begin{eqnarray}
q_{int,0} =  \frac{  1-3b^2}{2}.\label{qintlim1}
\end{eqnarray}
Therefore, we obtain that Eq. (\ref{qintlim1}) and (\ref{qintlim2}) is a constant  which depend onlys on the value of the interaction term $b^2$. Considering that $b^2=0.025$, we obtain that $q_{int,0} \approx 0.4625$, i.e. a value lower than what obtained for the non interacting case.\\
Instead, in the limiting case of $t\rightarrow \infty$, we obtain that:
\begin{eqnarray}
q_{int,late} =  \frac{1}{2}\left[ 1- \frac{  \left(3n+m-2\right)}{n  }   \right]   ,\label{qintlim2}
\end{eqnarray}
which is the same result of the non interacting case.\\
Then, we obtain that Eq. (\ref{qintlim2}) is a constant quantity which depends only on the values of $m$ and $n$.\\
At present time, we obtain that:
\begin{eqnarray}
q_{non,present,1}   &\approx&  -0.509, \label{decelnon1lim1prin}\\
q_{non,present,2}   &\approx& -0.407, \label{decelnon1lim2prin}\\
q_{non,present,3}   &\approx& -0.377,  \label{decelnon1lim3prin}
\end{eqnarray}
therefore we obtain values of the deceleration parameter $q$ indicating an accelerated Universe since they are negative.\\
Comparing the results of the non interacting and the interacting case, we can conclude that the presence of interaction between DE and DM affect the result obtained.\\
Following the same procedure of the non interacting case, we obtain following conditions about the parameters involved:
\begin{eqnarray}
m\left(l_2-1\right)-3 \left(1-b^2\right) n+2 &=&0,\label{rambone2}\\
2+m\left( l_1 -1 \right)&=&0,\label{rambone3}\\
m(l_2 -1) -2 &=&0,\label{rambone4}
\end{eqnarray}
Combining the results of Eqs. (\ref{rambone2}) and (\ref{rambone3}), we obtain also the following condition:
\begin{eqnarray}
m\left(l_2-1\right)-3 \left(1-b^2\right) n-m\left( l_1 -1 \right) &=&0,\label{rambone2lu}
\end{eqnarray}
which is equivalent to the following relation:
\begin{eqnarray}
m\left(l_2-l_1\right)-3 \left(1-b^2\right) n &=&0.\label{rambone2lulo}
\end{eqnarray}
Moreover, combining Eqs. (\ref{rambone2}) and (\ref{rambone4}), we obtain:
\begin{eqnarray}
2m\left(l_2-1\right)-3 \left(1-b^2\right) n &=&0.\label{rambone2new}
\end{eqnarray}
In the limiting case of $b^2=0$, i.e. for absence of interaction,  we recover the same results of the non interacting case.\\
\begin{table}[bht]\label{tablesura111}
\caption{Summary of the values of the Equation of State (EoS) parameter $w_D$ obtained using different observational cosmological data.}
\centering
\begin{small}
\begin{tabular}{|c|c|c|c|c|}
\hline $w_D$ & $ Observational$ $Schemes$ & $ References$\\
\hline$-1.13^{+0.24}_{-0.25}$ & Planck+WP+BAO & Ade  et al. (2013) \cite{planck} \\
\hline$-1.09\pm0.17$& Planck+WP+Union 2.1& Ade  et al. (2013) \cite{planck} \\
\hline$-1.13^{+0.13}_{-0.14}$ & Planck+WP+SNLS & Ade  et al. (2013) \cite{planck} \\
\hline$-1.24^{+0.18}_{-0.19}$ & WMAP+eCMB+BAO+$H_0$+SNe Ia & Ade  et al. (2013) \cite{planck} \\
\hline$-1.073^{+0.090}_{-0.089}$ & WMAP+eCMB+BAO+$H_0$  &  Hinshaw  et al. (2013) \cite{hins} \\
\hline$-1.084\pm0.063$& WMAP+eCMB+BAO+$H_0$+SNe Ia  & Hinshaw  et al. (2013) \cite{hins} \\
\hline
\end{tabular}
\end{small}
\end{table}
\par

\section{Statefinder Diagnostic}
The investigation and the study of some important  cosmological quantities (like for example the EoS parameter $w_D$, the Hubble parameter $H$  and the deceleration parameter $q$)  have attracted a lot of attention in modern Cosmology. Anyway, it is well known in Cosmology that different DE models usually lead to a positive value of the Hubble parameter $H$ and a negative value of deceleration parameter $q$ (i.e. they lead to $H > 0$ and $q < 0$) at the present day epoch of the Universe, i.e. for $t=t_0$, for this reason we can conclude that the Hubble and the deceleration parameters $H$ and $q$ can not effectively discriminate between the various DE models taken into account. For this reason, a higher order of  derivatives with respect to the cosmic time $t$ of the scale factor $a\left( t \right)$ must be taken into account and required if we want to have a better and deeper comprehension of the DE model taken into account. For this purpose, Sahni et al. \cite{sah} and Alam et al. \cite{alam}, considering the third derivative with respect to the cosmic time $t$ of the scale factor $a(t)$, recently introduced the statefinder pair $\left\{r,s\right\}$ with the aim to remove the degeneracy of $H$ and $q$ at the present epoch of the Universe.\\
The general expressions of the statefinder parameters $r$ and $s$ as functions of the total energy density and the total pressure are given, respectively, by the following general relations:
\begin{eqnarray}
r &=&  1+ \frac{9}{2}\left(\frac{ \rho_{tot} + p_{tot}   }{\rho_{tot} }\right)\left(\frac{\dot{p}_{tot} } {\dot{\rho}_{tot} } \right), \label{rgen}\\
s &=& \left(\frac{ \rho_{tot}  + p_{tot}   }{p_{tot} }\right)\left(\frac{\dot{p}_{tot} } {\dot{\rho}_{tot} } \right), \label{sgen}
\end{eqnarray}
or, equivalently, using the facts that $p_{tot} =p_D$ and $\rho_{tot}  = \rho_D + \rho_m$, by:
\begin{eqnarray}
r &=& 1+ \frac{9}{2}\left(\frac{ \rho_D + \rho_m + p_D  }{\rho_D + \rho_m}\right)\left(\frac{\dot{p}_D}{\dot{\rho}_D +\dot{\rho}_m}\right),      \label{rgen1}\\
s &=& \left(\frac{ \rho_D + \rho_m + p_D  }{p_D}\right)\left(\frac{\dot{p}_D}{\dot{\rho}_D +\dot{\rho}_m}\right).       \label{sgen1}
\end{eqnarray}
The expressions of the statefinder parameters $r$ and $s$ are also given, respectively, by the following relations:
\begin{eqnarray}
r &=& \frac{\dot{\ddot{a}}} {aH^3},  \label{r1}\\
s &=&   \frac{r -1}{3\left(q-1/2\right)},   \label{s1}
\end{eqnarray}
where $q$ represents the deceleration parameter, which has been already studied in the previous Section.\\
An alternative way to write the statefinder parameters $r$ and $s$ using the Hubble parameter and its time derivatives is the following one:
\begin{eqnarray}
r&=& 1 + 3\left(\frac{\dot{H}} {H^2}\right)+ \frac{\ddot{H}} {H^3},  \label{r2}\\
s&=& -\frac{3H\dot{H}+\ddot{H}} {3H\left( 2\dot{H}+3H^2  \right)} =  -\frac{3\dot{H}+\ddot{H}/H}{3\left( 2\dot{H}+3H^2  \right)}. \label{s2}
\end{eqnarray}
One of the most important properties of the statefinder parameters $r$ and $s$ is that the point with coordinate corresponding to $\left\{r, s\right\} = \left\{1, 0\right\}$ in the $r-s$ plane indicates the point corresponding to the flat $\Lambda$CDM model \cite{huang}. Therefore, we have that departures of given DE models from this fixed point are good ways to establish the distance of these models from the flat $\Lambda$CDM model. \\
Moreover, we must underline here that, in the $r - s$ plane, a positive value of the statefinder parameter $s$ (i.e.,  $s > 0$) indicates a quintessence-like model of DE while a negative value of the statefinder parameter $s$ (i.e. $s < 0$) indicates a phantom-like model of DE. Furthermore, an evolution from phantom to quintessence (or the inverse) is obtained when the  point with coordinates  $\left\{r, s\right\} = \left\{1, 0\right\}$ in the $r - s $ plane is crossed \cite{wu1}.\\
Different models, like braneworld, the Cosmological Constant $\Lambda_{CC}$, Chaplygin gas and quintessence, were well investigated in the paper of Alam et al. \cite{alam} using the statefinder diagnostic obtaining that the statefinder pair could effectively differentiate between these different models. An investigation on the statefinder parameters with the purpose of differentiate between DE and modified gravity models was carried out in the paper of Wang et al. \cite{wang}. The statefinder diagnostics for the $f\left(T\right)$ modified gravity model has been studied in the paper of Wu $\&$ Yu \cite{wu1}. \\
We now want to study the statefinder diagnostic for the model we are taking into account for both non interacting and interacting Dark Sectors.

\subsection{Non Interacting Case}
\begin{figure}[ht]
\centering\includegraphics[width=8cm]{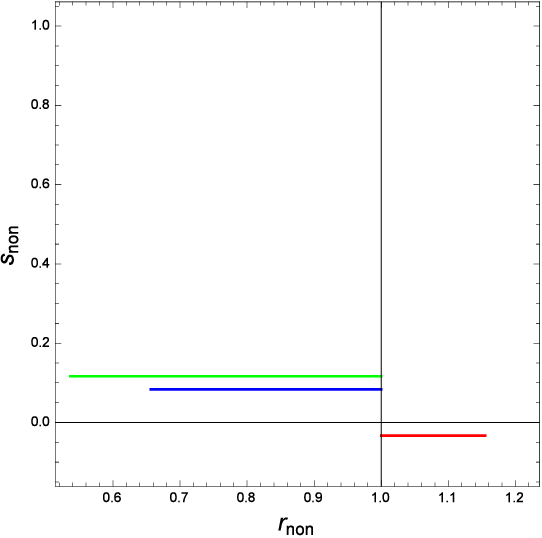}
\caption{The $\{r-s\}$ trajectories for the non interacting case.} \label{statenon}
\end{figure}

We start studying the behavior of the statefinder parameters $r$ and $s$ for the non interacting case.\\
Using the expressions of $\rho_D$, $p_D$ and $\rho_m$ given, respectively, in Eqs. (\ref{rhod2}) , (\ref{15}) and (\ref{rhom}), we obtain the following expressions for the statefinder parameters $r_{non} $ and $s_{non} $ for the non interacting case:
\begin{eqnarray}
r_{non} &=&  1+\frac{3 (m-2) (m+3 n-2) t^{m+3 n} \left[2 \alpha +n (n \beta -\epsilon )\right] a_0^3 \phi _0}{2 n^2 \left\{ \rho_{m0}t^2+3 t^{m+3 n}  \left[2 \alpha +n (n \beta -\epsilon )\right] a_0^3 \phi _0\right\}}  \nonumber\\
&=&  1+\frac{3 (m-2) (m+3 n-2) \left[2 \alpha +n (n \beta -\epsilon )\right] a_0^3 \phi _0}{2 n^2 \left\{ \rho_{m0}t^{2-m-3n}+3  \left[2 \alpha +n (n \beta -\epsilon )\right] a_0^3 \phi _0\right\}},  \label{rnon} \\
s_{non} &=& \frac{2-m}{3 n}.\label{snon}
\end{eqnarray}
Therefore, we obtain that the expression of the statefinder parameter $s$ obtained in Eq. (\ref{snon}) is a constant depending on the values of the parameters $m$ and $n$ only.\\
In Figure \ref{statenon}, we plotted the statefinder trajectories for the non interacting case using the results of Eqs. (\ref{rnon}) and (\ref{snon}).\\
We can now make some considerations about the values assumed by $r_{non}$ and $s_{non}$.\\
Singularities in the expressions of $r_{non}$ and $s_{non}$ are avoided since we cannot have $n=0$ since we obtained at the beginning of the paper that we must have $n>0$ in order to have an accelerated Universe.\\
For $t\rightarrow 0$, we obtain that $r\rightarrow 1$ while the expression of $s$ is given by Eq. (\ref{snon}). For the three cases considered, we obtain, therefore, the following set of values of the statefinder parameters:
\begin{eqnarray}
\left\{r_1,s_1   \right\}_{non,0} &\approx&  \left\{ 1, -0.033  \right\}, \\
\left\{r_2,s_2   \right\}_{non,0}  &\approx&  \left\{ 1,0.083   \right\},\\
\left\{r_3,s_3   \right\}_{non,0}  &\approx&  \left\{  1, 0.117 \right\}.
\end{eqnarray}
Therefore, for $t\rightarrow 0$, we obtain for the first two cases considered, i.e. for $m=2.2$ and $m=1.5$, values of the statefinder parameters which slightly differ from the value corresponding to the flat $\Lambda$CDM model: the departure from the flat $\Lambda$CDM model is due to the value assumed by $s$. Moreover, we obtain, for the case corresponding to $m=2.2$, a quintessence-like model since we obtain a negative value of  $s$ (i.e. $s<0$) while for the other two models considered we obtain a phantom-like model since we derive positive values of $s$ (i.e. $ s>0$).\\
At late times, i.e. for $t\rightarrow \infty$, we obtain that the expressions of the statefinder parameters are given by the following relations:
\begin{eqnarray}
r_{non,late} &=&  1+\frac{ (m-2) (m+3 n-2)  }{2 n^2 },   \label{rnonlim}\\
s_{non,late} &=& s_{non}= \frac{2-m}{3 n}.\label{snonlim}
\end{eqnarray}
Combining the results of Eqs. (\ref{rnonlim}) and (\ref{snonlim}), we can also write the following relation between $r_{non}$ and $s_{non}$:
\begin{eqnarray}
r_{non,late} &=&  1+\left(\frac{ 9  }{2 }\right)s_{non,late}\left( s_{non,late}-1 \right).   \label{rnonlim2}
\end{eqnarray}
For the three cases considered, i.e. $m=2.2$, $m=1.5$ and $m=1.3$ (with $n=2$ for all cases), we obtain therefore the following values of the statefinder pair:
\begin{eqnarray}
\left\{r_1,s_1   \right\}_{non,late} &\approx& \left\{1.155,-0.033 \right\}, \\
\left\{r_2,s_2   \right\}_{non,late}&\approx& \left\{0.656,0.083    \right\},\\
\left\{r_3,s_3   \right\}_{non,late} &\approx& \left\{0.536,0.117    \right\}.
\end{eqnarray}
For all the three cases considered, we obtain points which differs from that of the $\Lambda$CDM model, Moreover, for $m=2.2$ we obtain a quintessence-like model since $s$ is negative, while for the other two models we have a phantom-like model since we obtain that $s$ assumes positive values.\\
At present time, we obtain the following set of values for the three different cases we consider:
\begin{eqnarray}
\left\{r_1,s_1   \right\}_{non,present} &\approx&  \left\{ 1.099, -0.033  \right\}, \\
\left\{r_2,s_2   \right\}_{non,present}  &\approx&  \left\{ 0.777,0.083   \right\},\\
\left\{r_3,s_3   \right\}_{non,present}  &\approx&  \left\{  0.698, 0.117 \right\}.
\end{eqnarray}
Therefore, we obtain for the case corresponding to $m=2.2$, values of the statefinder parameters which slightly differ from that of the $\Lambda$CDM model, while for the other two cases we obtain values which considerably differs. Moreover, for the case corresponding to $m=2.2$, we obtain a quintessence-like model since $s$ is negative, while for the other two models we have a phantom-like model since we obtain that $s$ assumes positive values.\\
For $m+3n-2=0$, which means $m=2-3n$, we obtain that $r_{non}=1$. Moreover, inserting $m=2-3n$ in Eq. (\ref{snonlim}), we obtain that $s_{non}$ assumes the value of 1 too. Therefore, we obtain a value of the pair $\left\{r,s\right\} $ which differs from the one corresponding to the flat $\Lambda$CDM one since we have that $s_{non}\neq 0$. Moreover, since we have $s>0$, it means we are dealing with a quintessence-like model.\\
For the limiting case corresponding to $m=2$, we obtain $r_{non}=1$ and $s_{non}=0$ independently on the values of the other parameters.

\subsection{Interacting Case}
We now extend the results obtained in the previous subsection considering the presence of interaction between the Dark Sectors. \\
We always use the general expressions for $r$ and $s$ given in Eqs. (\ref{rgen1}) and (\ref{sgen1}). \\
We must underline here that for the interacting case the expression of $\rho_{m,int}$ is given by:
\begin{eqnarray}
\rho_{m,int} &=&\rho_{m0}  a^{3\left(1-b^2\right)}\nonumber\\
 &=& \rho_{m0}a_0^{-3\left(1-b^2\right)}t^{-3n\left(1-b^2\right)}   \label{romint} ,
\end{eqnarray}
where we used the expression of the scale factor $a(t)$ defined in Eq. (\ref{12}). \\
Therefore, using the expression of the pressure of DE and the energy density of DE for the interacting case, the expression of $\rho_m$ given in Eq. (\ref{romint}) and the expression of $q_{int}$ given in Eq. (\ref{qint!!}), we obtain the following expressions for the statefinde parameters $r_{int}$ and $s_{int}$ for the interacting case:
\begin{eqnarray}
r_{int} &=& 1+\frac{A_1}{B_1},  \label{mancinelli1} \\
s_{int} &=& \frac{A_2}{B_2},   \label{mancinelli2}
\end{eqnarray}
where the terms $A_1$ and $B_1$ are defined as follows:
\begin{eqnarray}
A_1  &=& 3 \left\{3 b^2 \left(b^2-1\right) n^2 t^{2+3 b^2 n} \rho _{m0}\nonumber \right.\\
&&\left.+(m-2) (m+3 n-2) t^{m+3 n} \left[2 \alpha +n (n \beta -\epsilon)\right] a_0^3 \phi _0\right\},  \\
B_1 &=& 2 n^2 \left\{t^{2+3 b^2 n} \rho _{m0}+3 t^{m+3 n} \left[2 \alpha +n (n \beta -\epsilon )\right] a_0^3 \phi _0\right\},
\end{eqnarray}
while the terms $A_2$ and $B_2$ are defined as follows:
\begin{eqnarray}
A_2 &=&t^{-\left(2+3n \right)} \left\{-\frac{3 b^2 \left(b^2-1\right) n t^{  -\left[1+3 \left(1-b^2\right) n\right] } \rho _{m0}} {a_0^3} \right. \nonumber \\
&&\left.-\frac{(m-2) (m+3n-2) t^{-3+m} \left[2 \alpha +n (n \beta -\epsilon )\right] \phi _0}{n}\right\}\times \nonumber \\
&&\left\{\left(b^2-1\right) n t^{2+3 b^2 n} \rho _{m0}+(m-2) t^{m+3 n} \left[2\alpha +n (n \beta -\epsilon )\right] a_0^3 \phi _0\right\} \\
B_2 &=& n a_0^3 \left\{\frac{3 \left(b^2-1\right) n t^{-\left[1+3 \left(1-b^2\right) n\right]} \rho_{m0}} {a_0^3}+3 \left(m-2\right) t^{-3+m} \left[2 \alpha +n \left(n \beta -\epsilon \right)\right] \phi _0\right\} \times \nonumber \\
&&\left\{\frac{b^2 t^{3 \left(b^2-1\right) n} \rho _{m0}} {a_0^3}+\frac{(m+3n-2) t^{-2+m} \left[2 \alpha +n (n \beta -\epsilon )\right] \phi _0}{n}\right\}.
\end{eqnarray}
We have that in the limiting case of $b^2=0$, i.e. in absence of interaction, we recover the same results of the non interacting case.\\
In the limiting case of $m=2$, we obtain the following expressions for the terms $A_1$, $A_2$, $B_1$ and $B_2$:
\begin{eqnarray}
A_1  &=& 3 \left[3 b^2 \left(b^2-1\right) n^2 t^{2+3 b^2 n} \rho _{m0} \right],  \\
A_2 &=&t^{-\left(2+3n \right)} \left\{-\frac{3 b^2 \left(b^2-1\right) n t^{  -\left[1+3 \left(1-b^2\right) n\right] } \rho _{m0}} {a_0^3} \right\}\times \nonumber \\
&&\left[\left(b^2-1\right) n t^{2+3 b^2 n} \rho _{m0}\right], \\
B_1 &=& 2 n^2 \left\{t^{2+3 b^2 n} \rho _{m0}+3 t^{2+3 n} \left[2 \alpha +n (n \beta -\epsilon )\right] a_0^3 \phi _0\right\}, \\
B_2 &=& n a_0^3 \left\{\frac{3 \left(b^2-1\right) n t^{-\left[1+3 \left(1-b^2\right) n\right]} \rho_{m0}} {a_0^3}\right\} \times \nonumber \\
&&\left\{\frac{b^2 t^{3 \left(b^2-1\right) n} \rho _{m0}} {a_0^3}+ 3\left[2 \alpha +n (n \beta -\epsilon )\right] \phi _0\right\}.
\end{eqnarray}
In the limiting case of $b^2=0$, we recover the same results of the non interacting case.\\
For $\rightarrow 0$, we obtain that the expression of $r_{int}$ reduces to:
\begin{eqnarray}
r_{int,0} = 1 +\left(\frac{9}{2}\right)b^2\left(b^2-1\right),
\end{eqnarray}
while $s_{int,0}$ is given by:
\begin{eqnarray}
s_{int,0} =1-b^2.
\end{eqnarray}
Therefore, we obtain values of the statefinder parameters depending only on the value of the interaction parameter $b^2$.\\
Considering the value $b^2=0.025$, we obtain the following values for the statefinder pair:
\begin{eqnarray}
\left\{r_{int,0},s_{int,0}\right\} \approx \left\{0.890,0.975   \right\}. \label{vera}
\end{eqnarray}
The result of Eq. (\ref{vera}) is valid for all the three cases considered.\\
Therefore, for $t\rightarrow 0$, for the interacting case, we obtain a point in the $r-s$ plane which considerably differs from the point corresponding to the flat $\Lambda$CDM model. Moreover, since we obtain a positive value of $s$, we conclude that we deal with a quintessence-like model.\\
Moreover, for the limiting case of $t\rightarrow \infty$, we recover the same result obtained for the non interacting case.\\
\begin{figure}[ht]
\centering\includegraphics[width=8cm]{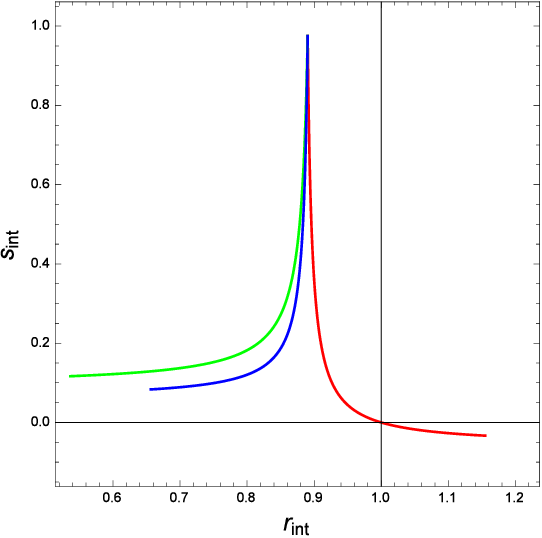}
\caption{The $\{r-s\}$ trajectories for the interacting case.} \label{stateint}
\end{figure}
At present time, we obtain the following set of values for the statefinder parameters:
\begin{eqnarray}
\left\{r_{1}, s_{1} \right\}_{int,present}&\approx& \left\{1.060, -0.020\right\},\\
\left\{r_{2},s_{2}  \right\}_{int,present}&\approx& \left\{0.738,0.096\right\},\\
\left\{r_{3}, s_{3} \right\}_{int,present}&\approx& \left\{0.659,  0.129\right\}.
\end{eqnarray}
For the case corresponding to $m=2.2$, we obtain a value of the statefidner pair which is quite close to that of the $\Lambda$CDM model, while for the other two cases we obtain values which considerably differ from it. Moreover, for the case corresponding to $m=2.2$, we obtain a quintessence-like model since $s$ is negative, while for the other two models we have a phantom-like model since we obtain that $s$ assumes positive values.\\
In Figure \ref{stateint}, we have studied the statefinder trajectories for the three cases we considered in this paper and we have observed that the flat $\Lambda$CDM point is attainable. Furthermore, the trajectory can go beyond that fixed point $\{r=1,s=0\}$  and it can reach the quadrant $r>1,s<0$ for the case with $m=2.2$ only.

\section{Cosmographic Parameters}
In this Section, we obtain some important cosmological information about the energy density model we are considering using the properties of the cosmographic parameters.\\
Standard candles (like for example SNe Ia) can be considered like powerful instruments in present day Cosmology since they can be safely used in order to reconstruct the Hubble diagram, i.e. the redshift-distance relation up to high redshifts $z$. It is quite common to constrain a particular model which is wanted to be studied against the data in order to check its validity and to constraint the values of its free parameters. Anyway, this kind of approach is highly model-dependent, for this reason there are still some doubts in the scientific community on the validity and reliability of the constraints on the derived cosmological quantities.\\
In order to avoid this kind of problem, it is possible to use the properties and features of cosmography, i.e. expanding the scale factor $a\left( t \right)$ in Taylor series with respect to the cosmic time $t$. This type of expansion of the scale factor leads to a distance-redshift relation which results to be totally model independent since it results to be independent on the particular form of the solution of the cosmic equations. Cosmography can be also considered as a milestone in the study of the main features of the dynamics of the Universe, which any theoretical model considered has to take into account and also to satisfy. We can now introduce the following five quantities \cite{cosmo1,cosmo2}:
\begin{eqnarray}
q &=& -\left(\frac{\ddot{a}} {a}\right)H^{-2}=  -\frac{a^{\left(2\right)}a}{\dot{a}^2}  \label{par1}, \\
j &=& \left(\frac{1}{a}\right)\left(\frac{d^3a}{dt^3}\right)H^{-3} = \frac{a^{\left(3\right)}a^2}{\dot{a}^3}  \label{par2}, \\
s_{cosmo} &=&\left( \frac{1}{a}\right)\left(\frac{d^4a}{dt^4}\right)H^{-4}= -\frac{a^{\left(4\right)}a^3}{\dot{a}^4}  \label{par3}, \\
l &=&   \left(\frac{1}{a}\right)\left(\frac{d^5a}{dt^5}\right)H^{-5}= -\frac{a^{\left(5\right)}a^4}{\dot{a}^5}, \label{par4}\\
m &=&   \left(\frac{1}{a}\right)\left(\frac{d^6a}{dt^6}\right)H^{-6}= -\frac{a^{\left(6\right)}a^5}{\dot{a}^6}. \label{par5}
\end{eqnarray}
In order to avoid ambiguities, we indicate the third quantity (which will be described later on) with $s_{cosmo}$ in order it is not confused with the statefinder parameter $s$ discussed in the previous Section.\\
In general, we have that the $i$-th parameter $x^i$ can be obtained thanks to the following expression:
\begin{eqnarray}
x^{i} &=& \left( -1  \right)^{i+1}\left(\frac{1}{H^{i}}\right) \frac{a^{\left(i\right)}} {a}= \left( -1  \right)^{i+1} \frac{a^{\left(i\right)}a^{i-1}} {\dot{a}^{i+1}},
\end{eqnarray}
where the index $i$, when in parenthesis, indicates the order of the derivative with respect to the cosmic time $t$ while, when not in parenthesis, indicates the power law index of the corresponding quantity.\\
The quantities given in Eqs. (\ref{par1}), (\ref{par2}), (\ref{par3}), (\ref{par4}) and (\ref{par5}) are known, respectively, as deceleration, jerk, snap, lerk and max-out parameters.   We have already derived and studied the expression of the deceleration parameter $q$ for both non interacting and non interacting Dark Sectors in the previous Section. \\
The present-day values of the cosmographic parameters, denoted with the subscript 0,  can be used in order to characterize the evolutionary status of the Universe. For example, a negative value of  $q_0$ indicates an accelerated expansion of the Universe, while the value of $j_0$ allows us to discriminate among different accelerating models. Moreover, $j_0$ can be seen as a parameter which indicates the status of the variation of the acceleration of the Universe; we also know that a positive value of $j_0$ implies that the deceleration parameter $q$ changes its sign as the Universe expands. In some recent works, some constrains about the values of the cosmographic snap and lerk parameters have been derived.   For example, Capozziello $\&$ Izzo \cite{capoizzo1}  obtained that $s_0 = 8.32 \pm 12.16$, while John \cite{capoizzo2,capoizzo3} has derived that $s_{cosmo,0}=36.5 \pm 52.9$ and $l_0=142.7 \pm 320$. As we can clearly see from the values obtained in these works, the errors associated with the values derived for the snap and lerk cosmographic parameters  are of the order of 200$\%$, therefore for future more precise comparisons between cosmological constraints of $s$ and $l$ and the values obtained from theoretical models, it will be useful to have better constraints with more accurate errors. Instead, in the work of Aviles et al. \cite{orlando2}, some constrains about the value of $m$ have been obtained. Authors found that $m_0 = 71.93^{+382.17}_{-316.76}$ using Union 2 + HST + $H\left(z\right)$ data. Moreover, using the same set of data, they derived $j_0=-0.117^{+3.612}_{-1.257}$, $s_{cosmo,0} = -7.71^{+14.77}_{-7.83}$ and $l_0=8.55^{+23.39}_{-27.86}$.\\
Using the definitions given in Eqs. (\ref{par1}), (\ref{par2}), (\ref{par3}),   (\ref{par4}) and  (\ref{par5}),   we can easily obtain the sixth order Taylor expansion of the scale factor $a\left( t \right)$ as follows:
\begin{eqnarray}
\frac{a\left(t\right)}{a\left(t_0\right)} &=& 1+H_0 \left( t -t_0 \right) - \left(\frac{q_0}{2!}\right)H_0^2 \left( t -t_0 \right)^2 +  \left(\frac{j_0}{3!}\right)H_0^3 \left( t -t_0 \right)^3  \nonumber \\
 &&+\left(\frac{s_{cosmo,0}}{4!}\right)H_0^4 \left( t -t_0 \right)^4   +    \left(\frac{l_0}{5!}\right)H_0^5 \left( t -t_0 \right)^5 \nonumber \\
&&+    \left(\frac{m_0}{6!}\right)H_0^6 \left( t -t_0 \right)^6  +O \left[ \left( t-t_0  \right)^7 \right], \label{exp}
\end{eqnarray}
where $t_0 $ represents the present day age of the Universe while $H_0$ indicates the present day value of the Hubble parameter $H$.\\
The deceleration parameter $q$ has been already introduced and studied in Section 2. \\
The jerk parameter $j$ is also another name of the statefinder parameter $r$ we have previously studied  and it represents a natural next step beyond the Hubble parameter $H$ and the deceleration parameter $q$ \cite{alam,arab} since it involves derivatives of the scale factor $a(t)$ with respect to the cosmic time higher than $H$ and $q$.\\
The snap parameter $s_{cosmo}$, which depends on the fourth time derivative of the scale factor $a\left(t \right)$, is also sometimes called kerk parameter and it has been well discussed in the works of Dabrowski \cite{dabro}  Dunajski $\&$ Gibbons \cite{duna} and Arabsalmania $\&$ Sahni \cite{arab}.\\
The lerk parameter $l$ depends on the fifth time derivative of scale factor. More information can be found in the paper of Dabrowski \cite{dabro}.\\
The $m$ parameter, known also as max-out parameter, was considered and studied, for example, in the work of Dunsby $\&$ Luongo \cite{orlando1} and Aviles et al. \cite{orlando2}.  \\
Some relations involving the first time derivatives of the cosmographic parameters and the cosmographic parameters themselves are the following ones:
\begin{eqnarray}
\frac{dq}{dt} &=& -H \left( j -2q^2 -q  \right),  \\
\frac{dj}{dt} &=&H \left[s_{cosmo}+j\left( 2+3q  \right)   \right], \\
\frac{ds_{cosmo}}{dt} &=&H \left[ l+s_{cosmo} \left(3+4q \right)  \right],  \\
\frac{dl}{dt} &=& H \left[m+l\left(4+5q \right)   \right].
\end{eqnarray}
Equivalent expressions involving the cosmographic parameters and the time derivatives of the Hubble parameter $H$ (i.e. the Hubble rate) are the following ones:
\begin{eqnarray}
\frac{d^2H}{dt^2} &=& H^3 \left( j + 3q + 2  \right), \label{}\\
\frac{d^3H}{dt^3} &=& H^4 \left[ s_{cosmo} -4j -3q\left( q + 4 \right) -6  \right], \label{}\\
\frac{d^4H}{dt^4} &=& H^5 \left[ l - 5s_{cosmo} + 10\left(q+2 \right)j + 30 \left(q+2 \right)q +24  \right], \label{}\\
\frac{d^5H}{dt^5} &=& H^6 \left\{ m - 10j^2 -120j\left(q+1 \right) \right.\nonumber \\
&&\left. -3\left[2l + 5\left(24q + 18q^2 + 2q^3-2s_{cosmo}-qs_{cosmo} +8    \right)   \right]   \right\},
\end{eqnarray}
from which we can easily derive the following relations for $j$, $s_{cosmo}$, $l$ and $m$:
\begin{eqnarray}
j &=& \frac{\ddot{H} }{H^3} - 3q -2,\label{}\\
s_{cosmo} &=& \left(\frac{1}{H^4}\right)\frac{d^3H}{dt^3} +4j +3q\left( q + 4 \right) +6, \label{}\\
l &=& \left(\frac{1}{H^5}\right)\frac{d^4H}{dt^4} + 5s_{cosmo} - 10\left(q+2 \right)j - 30 \left(q+2 \right)q -24, \label{}\\
m&=& \left(\frac{1}{H^6}\right)\frac{d^5H}{dt^5} + 10j^2 + 120j\left( q+1  \right) \nonumber \\
&&+ 3 \left[ 2l + 5\left(24q + 18q^2 + 2q^3-2s_{cosmo}-qs_{cosmo} +8    \right)  \right].\label{}
\end{eqnarray}
We must underline that we have already derived the expression of $q$ for both non interacting and interacting cases in Eqs. (\ref{decelnon1}) and (\ref{qint!!}), respectively. \\
We now want to derive information about the cosmographic parameters above defined for the model we are studying and for both non interacting and later on interacting DE and DM.

\subsection{Non Interacting Case}
\begin{figure}[ht]
\centering\includegraphics[width=8cm]{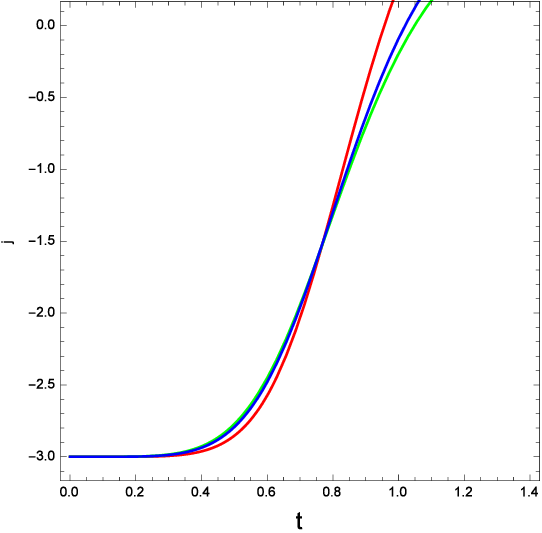}
\caption{Plot of the cosmographic parameter $j_{non}$ obtained in Eq. (\ref{jnoncosmo}) as function of the cosmic time $t$. } \label{jcosmo}
\end{figure}

\begin{figure}[ht]
\centering\includegraphics[width=8cm]{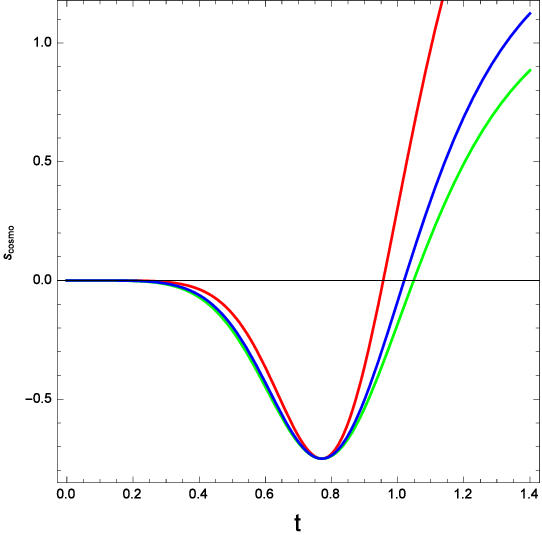}
\caption{ Plot of the cosmographic parameter $s_{cosmo,non}$ obtained in Eq. (\ref{snoncosmo}) as function of the cosmic time $t$. } \label{scosmo}
\end{figure}

\begin{figure}[ht]
\centering\includegraphics[width=8cm]{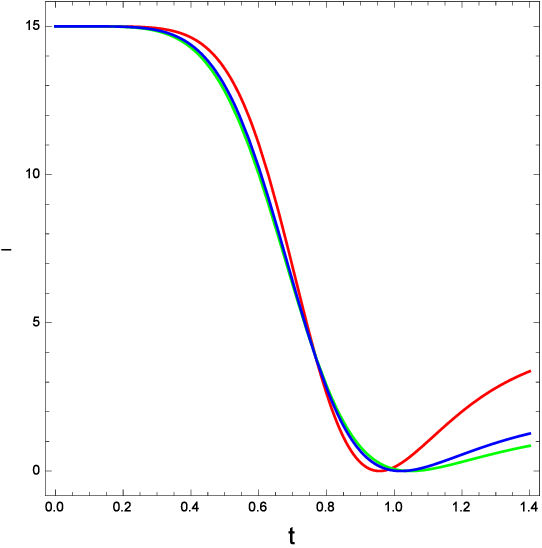}
\caption{Plot of the cosmographic parameter $l_{non}$ obtained in Eq. (\ref{lnoncosmo}) as function of the cosmic time $t$.  } \label{lcosmo}
\end{figure}

\begin{figure}[ht]
\centering\includegraphics[width=8cm]{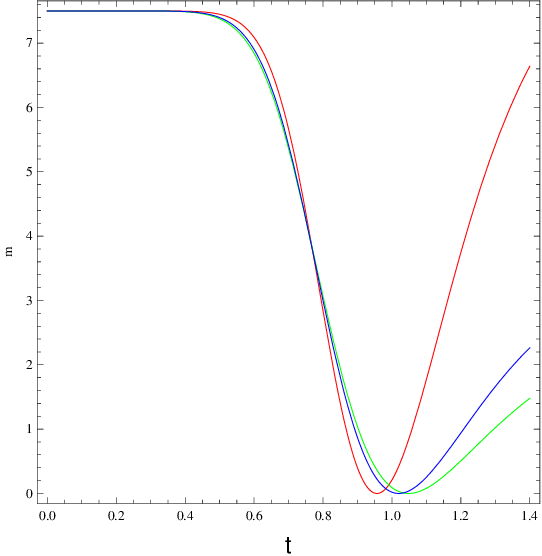}
\caption{Plot of the cosmographic parameter $m_{non}$ obtained in Eq. (\ref{mnoncosmo}) as function of the cosmic time $t$.  } \label{mcosmo}
\end{figure}

We start calculating the expressions of the cosmographic parameters for the non interacting case.\\
Using the expression of the scale factor defined in  in Eq. (\ref{12}), we obtain the following relation involving the Hubble parameter $H$ and its time derivatives:
\begin{eqnarray}
\frac{\ddot{H} }{H^3} &=& \frac{2}{n^2},  \label{h1}\\
\frac{1}{H^4}\left(\frac{d^3H}{dt^3}\right) &=& -\frac{6}{n^3},  \label{h2}\\
\frac{1}{H^5}\left(\frac{d^4H}{dt^4}\right)  &=& \frac{24}{n^4}, \label{h3}\\
\frac{1}{H^6}\left(\frac{d^5H}{dt^5}\right)  &=&- \frac{120}{n^5}. \label{h4}
\end{eqnarray}
Therefore, using the results of Eqs. (\ref{h1}), (\ref{h2}), (\ref{h3}) and (\ref{h4}), we obtain the following expressions for the cosmographic parameters for the non interacting case:
\begin{eqnarray}
j_{non} &=& \frac{2}{n^2} - 3q_{non} -2, \label{jnoncosmo}\\
s_{cosmo,non} &=&  3q_{non}^2  -\frac{6}{n^3} + \frac{8}{n^2} -2, \label{snoncosmo}\\
l_{non} &=& 15q_{non}^2 +20q_{non}   +\frac{24}{n^4} -\frac{30}{n^3} -\frac{20q_{non}} {n^2} +6, \label{lnoncosmo}\\
m_{non}  &=&  -15 q_{non}^3  +30q_{non}   - \frac{120}{n^5}   +\frac{184}{n^4}+\frac{90q_{non}  }{n^3}-\frac{80}{n^2}-\frac{120q_{non}}{n^2}+16,  \label{mnoncosmo}
\end{eqnarray}
where $q_{non}$ has been obtained  in Eq. (\ref{decelnon1}). \\
In Figures \ref{jcosmo}, \ref{scosmo}, \ref{lcosmo} and \ref{mcosmo}, we plot the expressions of $j_{non}$, $s_{cosmo,non}$, $l_{non}$ and $m_{non}$ we have obtained, respectively,  in Eqs.  (\ref{jnoncosmo}), (\ref{snoncosmo}), (\ref{lnoncosmo})  and (\ref{mnoncosmo}).

For $t\rightarrow0$, we obtain the following expressions for $j_{non,0}$, $s_{cosmo,non,0}$, $l_{non,0}$ and $m_{non,0}$:
\begin{eqnarray}
j_{non,0} &=& \frac{2}{n^2} - 3q_{non,0} -2, \label{}\\
s_{cosmo,non,0} &=&  3q_{non,0}^2  -\frac{6}{n^3} + \frac{8}{n^2} -2, \label{}\\
l_{non,0} &=& 15q_{non,0}^2 +20q_{non,0}   +\frac{24}{n^4} -\frac{30}{n^3} -\frac{20q_{non,0}} {n^2} +6, \label{}\\
m_{non,0} &=& -15  q_{non,0}^3 + 30q_{non,0} - \frac{120}{n^5}  +\frac{184}{n^4}+\frac{90q_{non,0}}{n^3}-\frac{80}{n^2}-\frac{120q_{non,0}}{n^2}+16.
\end{eqnarray}
Considering that $q_{non,0}=\frac{1}{2}$ as we have derived before, we obtain the following expressions for $j_{non,0}$, $s_{cosmo,non,0}$, $l_{non,0}$ and $m_{non,0}$:
\begin{eqnarray}
j_{non,0} &=& \frac{2}{n^2} - \frac{7}{2}, \label{}\\
s_{cosmo,non,0} &=&  -\frac{5}{4}  -\frac{6}{n^3} + \frac{8}{n^2} , \label{}\\
l_{non,0} &=& \frac{79}{4}   +\frac{24}{n^4} -\frac{30}{n^3} -\frac{10} {n^2} , \label{}\\
m_{non,0} &=& \frac{233}{8} - \frac{120}{n^5}  +\frac{184}{n^4}+\frac{45}{n^3}-\frac{140}{n^2}.
\end{eqnarray}
Using the value of $n$ we have considered, i.e. $n=2$, we obtain the following values for $j_{non,0}$, $s_{cosmo,non,0}$, $l_{non,0}$ and $m_{non,0}$:
\begin{eqnarray}
j_{non,0} &\approx& -3, \label{trapa1}\\
s_{cosmo,non,0} &\approx&  0, \label{trapa2}\\
l_{non,0} &\approx& 15, \label{trapa3}\\
m_{non,0} &\approx& 7.5. \label{trapa4}
\end{eqnarray}
We can observe that the results obtained in Eqs. (\ref{trapa1}), (\ref{trapa2}), (\ref{trapa3}) and (\ref{trapa4}) are in agreement with the results obtained in Figures \ref{jcosmo}, \ref{scosmo}, \ref{lcosmo} and \ref{mcosmo}.\\
Therefore, we conclude that we obtain values of the cosmographic parameters which are in agreement with results found in other papers since the values we derived are between the errors bars of the values found in the relevant papers.\\
For far future, i.e. for $t\rightarrow \infty$, we have that $j_{non,late}$, $s_{cosmo,non,late}$, $l_{non,late}$ and $m_{non,late}$ are given, respectively, by the following relations:
\begin{eqnarray}
j_{non,late} &=& \frac{2}{n^2} - 3q_{non,late} -2, \label{}\\
s_{cosmo,non,late} &=&  3q_{non,late}^2  -\frac{6}{n^3} + \frac{8}{n^2} -2, \label{}\\
l_{non,late} &=& 15q_{non,late}^2 +20q_{non,late}   +\frac{24}{n^4} -\frac{30}{n^3} -\frac{20q_{non,late}} {n^2} +6, \label{}\\
m_{non,late} &=& -15  q_{non,late}^3 + 30q_{non,late} - \frac{120}{n^5}  +\frac{184}{n^4}\nonumber \\
&&+\frac{90q_{non,late}}{n^3}-\frac{80}{n^2}-\frac{120q_{non,late}}{n^2}+16.
\end{eqnarray}
Considering the expression of $q_{non,late}$    given in Eq. (\ref{decelnon1lim}), we can write the following expressions for $j_{non,late}$, $s_{cosmo,non,late}$, $l_{non,late}$ and $m_{non,late}$:
\begin{eqnarray}
j_{non,late} &=& \frac{2}{n^2} + 3\left[ 1+ \frac{  \left(m-2\right)}{2n  }    \right] -2 \nonumber \\
&=& \frac{2}{n^2} +\frac{  3\left(m-2\right)}{2n  } +1, \label{}\\
s_{cosmo,non,late} &=&  3\left[  1+ \frac{  \left(m-2\right)}{2n  }   \right]^2  -\frac{6}{n^3} + \frac{8}{n^2} -2, \label{}\\
l_{non,late} &=& 15\left[ 1+ \frac{  \left(m-2\right)}{2n  }   \right]^2 -20\left[  1+ \frac{  \left(m-2\right)}{2n  }  \right]  \nonumber \\
 &&+\frac{24}{n^4} -\frac{30}{n^3} +\frac{20} {n^2}\left[  1+ \frac{  \left(m-2\right)}{2n  }   \right] +6, \label{}\\
m_{non,late} &=& 15 \left[  1+ \frac{  \left(m-2\right)}{2n  } \right]^3  -30\left[ 1+ \frac{  \left(m-2\right)}{2n  }   \right] - \frac{120}{n^5}  \nonumber \\
&&+\frac{184}{n^4}-\frac{90}{n^3}\left[ 1+ \frac{  \left(m-2\right)}{2n  }   \right]-\frac{80}{n^2}+\frac{120}{n^2}\left[ 1+ \frac{  \left(m-2\right)}{2n  }   \right]+16.
\end{eqnarray}
Therefore, for the three cases considered for $m$ and $n$, i.e. $m=2.2$, $m=1.5$ and $m=1.3$ with $n=2$ for all the three cases, we obtain the following values for the cosmographic parameters:
\begin{eqnarray}
j_{non,late,1} &\approx& 1.650,\\
j_{non,late,2} &\approx& 1.125,\\
j_{non,late,3} &\approx& 0.975,
\end{eqnarray}

\begin{eqnarray}
s_{cosmo,non,late,1} &\approx& 2.557,\\
s_{cosmo,non,late,2} &\approx& 1.547,\\
s_{cosmo,non,late,3} &\approx& 1.292,
\end{eqnarray}
\begin{eqnarray}
l_{non,late,1} &\approx& 4.537,\\
l_{non,late,2} &\approx& 2.109,\\
l_{non,late,3} &\approx& 1.584,
\end{eqnarray}
\begin{eqnarray}
m_{non,late,1} &\approx& 9.302,\\
m_{non,late,2} &\approx& 3.955,\\
m_{non,late,3} &\approx& 2.891.
\end{eqnarray}
At present time, we obtain, for the three different cases we are considering, the following values for the cosmographic parameters we are studying:
\begin{eqnarray}
j_{non,present,1} &\approx& -0.0209,\\
j_{non,present,2} &\approx& -0.326,\\
j_{non,present,3} &\approx& -0.414,
\end{eqnarray}
\begin{eqnarray}
s_{cosmo,non,present,1} &\approx& -0.0208,\\
s_{cosmo,non,present,2} &\approx& -0.290,\\
s_{cosmo,non,present,3} &\approx& -0.357,
\end{eqnarray}
\begin{eqnarray}
l_{non,present,1} &\approx& 0.0007,\\
l_{non,present,2} &\approx& 0.177,\\
l_{non,present,3} &\approx& 0.286,
\end{eqnarray}
\begin{eqnarray}
m_{non,present,1} &\approx& 0.001,\\
m_{non,present,2} &\approx& 0.246,\\
m_{non,present,3} &\approx& 0.390.
\end{eqnarray}
The present day values of the cosmographic parameters we have derived are in agreement (or between the errors)  with the results obtained in some recent papers.

\subsection{Interacting Case}
\begin{figure}[ht]
\centering\includegraphics[width=8cm]{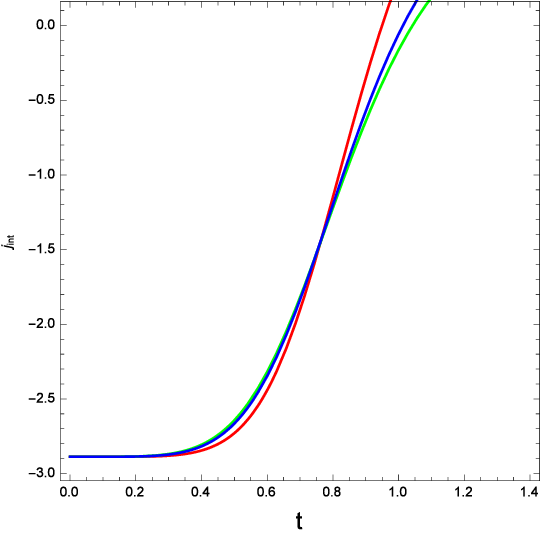}
\caption{ Plot of the cosmographic parameter $j_{int}$ obtained in Eq. (\ref{jintcosmo}) as function of the cosmic time $t$. } \label{jcosmoint}
\end{figure}

\begin{figure}[ht]
\centering\includegraphics[width=8cm]{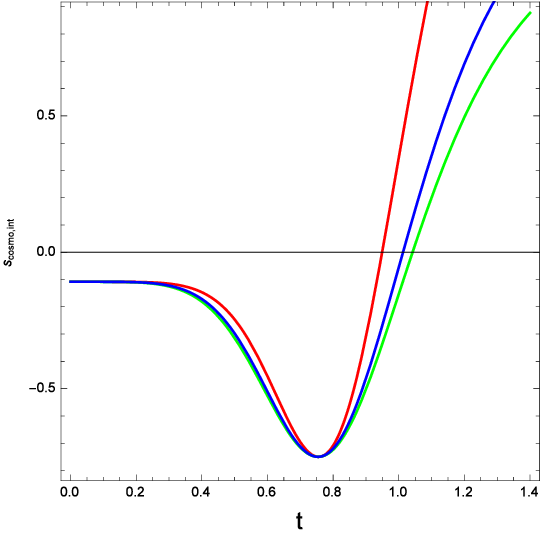}
\caption{ Plot  of the cosmographic parameter $s_{cosmo,int}$ obtained in Eq. (\ref{sintcosmo}) as function of the cosmic time $t$. } \label{scosmoint}
\end{figure}

\begin{figure}[ht]
\centering\includegraphics[width=8cm]{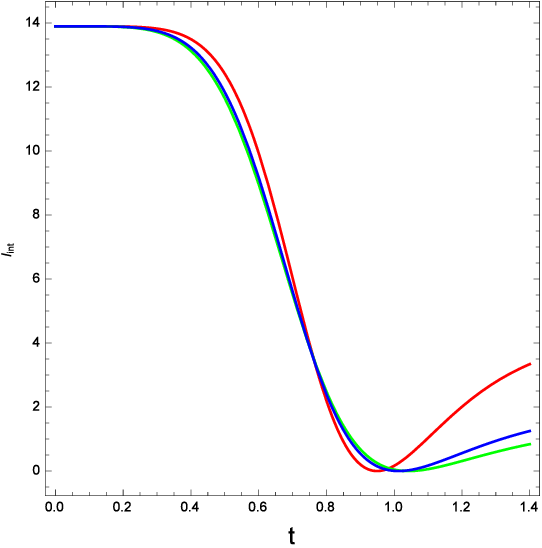}
\caption{Plot of the cosmographic parameter $l_{int}$ obtained in Eq. (\ref{lintcosmo}) as function of the cosmic time $t$.  } \label{lcosmoint}
\end{figure}

\begin{figure}[ht]
\centering\includegraphics[width=8cm]{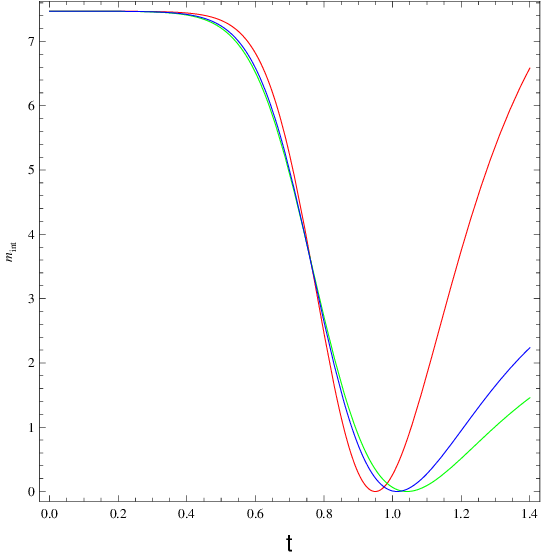}
\caption{Plot of the cosmographic parameter $m_{int}$ obtained in Eq. (\ref{mintcosmo}) as function of the cosmic time $t$.  } \label{mcosmoint}
\end{figure}

We now want to derive the expressions of the cosmographic parameters  considering the presence of  interaction between the Dark Sectors.\\
We still use the general expressions for $j$, $s_{cosmo}$, $l$ and $m$:
\begin{eqnarray}
j &=& \frac{\ddot{H} }{H^3} - 3q -2, \label{}\\
s_{cosmo} &=& \left(\frac{1}{H^4}\right)\frac{d^3H}{dt^3} +4j +3q\left( q + 4 \right) +6 , \label{}\\
l &=& \left(\frac{1}{H^5}\right)\frac{d^4H}{dt^4} + 5s_{cosmo} - 10\left(q+2 \right)j - 30 \left(q+2 \right)q -24 , \label{}\\
m &=& \left(\frac{1}{H^6}\right)\frac{d^5H}{dt^5} + 10j^2 + 120j\left( q+1  \right) \nonumber \\
&&+ 3 \left[ 2l + 5\left(24q + 18q^2 + 2q^3-2s_{cosmo}-qs_{cosmo} +8    \right)  \right].\label{}
\end{eqnarray}
We have already derived the expression of $q_{int}$ in Eq. (\ref{qint}).  Moreover, the expressions of $\frac{\ddot{H} }{H^3} $, $\left(\frac{1}{H^4}\right)\left(\frac{d^3H}{dt^3}\right) $, $\left(\frac{1}{H^5}\right)\left(\frac{d^4H}{dt^4}\right)$ and $\left(\frac{1}{H^6}\right)\left(\frac{d^5H}{dt^5}\right)$ are the same of Eqs. (\ref{h1}), (\ref{h2}), (\ref{h3}) and  (\ref{h4}).  Therefore, we obtain, for the interacting case, that $j_{int}$, $s_{cosmo,int}$, $l_{int}$ and $m_{int}$ are given by the following relations:
\begin{eqnarray}
j_{int} &=& \frac{2}{n^2} - 3q_{int} -2, \label{jintcosmo}\\
s_{cosmo,int} &=&  3q_{int}^2  -\frac{6}{n^3} + \frac{8}{n^2} -2, \label{sintcosmo}\\
l_{int} &+&  15q_{int}^2 +20q_{int}   +\frac{24}{n^4} -\frac{30}{n^3} -\frac{20q_{int}} {n^2} +6, \label{lintcosmo}\\
m_{int} &=&  -15 q_{int}^3 + 430q_{int} - \frac{120}{n^5}  +\frac{184}{n^4}+\frac{90q_{int}}{n^3}-\frac{80}{n^2}-\frac{120q_{int}}{n^2}+16, \label{mintcosmo}
\end{eqnarray}
where $q_{int}$ has been obtained  in Eq. (\ref{qint!!}). \\
In Figures \ref{jcosmoint}, \ref{scosmoint}, \ref{lcosmoint} and \ref{mcosmoint}, we plot the expressions of $j_{int}$, $s_{cosmo,int}$, $l_{int}$ and $m_{int}$ we have obtained, respectively,  in Eqs.  (\ref{jintcosmo}), (\ref{sintcosmo}), (\ref{lintcosmo})  and (\ref{mintcosmo}).

For $t\rightarrow0$, we obtain the following expressions for $j_{int,0}$, $s_{cosmo,int,0}$, $l_{int,0}$ and $m_{int,0}$:
\begin{eqnarray}
j_{int,0} &=& \frac{2}{n^2} - 3q_{int,0} -2, \label{maria1}\\
s_{cosmo,int,0} &=&  3q_{int,0}^2  -\frac{6}{n^3} + \frac{8}{n^2} -2, \label{maria2}\\
l_{int,0} &=&   15q_{int,0}^2 +20q_{int,0}   +\frac{24}{n^4} -\frac{30}{n^3} -\frac{20q_{int,0}} {n^2} +6, \label{maria3}\\
m_{int,0} &=&  -15 q_{int,0}^3 + 30q_{int,0} - \frac{120}{n^5}  +\frac{184}{n^4}+\frac{90q_{int,0}}{n^3}-\frac{80}{n^2}-\frac{120q_{int,0}}{n^2}+16, \label{maria4}
\end{eqnarray}
where $q_{int,0}$ has been obtained in Eq. (\ref{qintlim1}). \\
Inserting the expression of $q_{int,0}$ in Eqs. (\ref{maria1}), (\ref{maria2}), (\ref{maria3}) and (\ref{maria4}), we can write:
\begin{eqnarray}
j_{int,0} &=& \frac{2}{n^2} -\frac{  1-3b^2}{2} -2, \label{maria11}\\
s_{cosmo,int,0} &=&  3\left(\frac{  1-3b^2}{2}\right)^2  -\frac{6}{n^3} + \frac{8}{n^2} -2, \label{maria22}\\
l_{int,0} &=&   15\left(\frac{  1-3b^2}{2}\right)^2 +20\left(\frac{  1-3b^2}{2}\right)   +\frac{24}{n^4} -\frac{30}{n^3} -\frac{20} {n^2}\left(\frac{  1-3b^2}{2}\right) +6, \label{maria33}\\
m_{int,0} &=&  -15 \left(\frac{  1-3b^2}{2}\right)^3 + 30\left(\frac{  1-3b^2}{2}\right) - \frac{120}{n^5} \nonumber \\
 &&+\frac{184}{n^4}+\frac{90}{n^3}\left(\frac{  1-3b^2}{2}\right)-\frac{80}{n^2}-\frac{120}{n^2}\left(\frac{  1-3b^2}{2}\right)+16. \label{maria44}
\end{eqnarray}
Inserting the value of $n$ we considered, i.e. $n=2$, and considering $b^2=0.025$, we obtain the following values for the cosmographic parameters:
\begin{eqnarray}
j_{int,0} &\approx& -2.888, \label{maria111}\\
s_{cosmo,int,0} &\approx&  -0.108, \label{maria222}\\
l_{int,0} &\approx&  13.896, \label{maria333}\\
m_{int,0} &\approx&  7.469. \label{maria444}
\end{eqnarray}
At late times, we obtain the same results of the non interacting case.\\
At present time, we obtain the following values for the cosmographic parameters we considered in this paper:
\begin{eqnarray}
j_{int,present,1} &\approx& 0.027,\\
j_{int,present,2} &\approx& -0.280,\\
j_{int,present,3} &\approx& -0.369,
\end{eqnarray}
\begin{eqnarray}
s_{cosmo,int,present,1} &\approx& 0.027,\\
s_{cosmo,int,present,2} &\approx& -0.254,\\
s_{cosmo,int,present,3} &\approx& -0.323,
\end{eqnarray}
\begin{eqnarray}
l_{int,present,1} &\approx& 0.0012,\\
l_{int,present,2} &\approx& 0.130,\\
l_{int,present,3} &\approx& 0.227,
\end{eqnarray}
\begin{eqnarray}
m_{int,present,1} &\approx& 0.0018,\\
m_{int,present,2} &\approx& 0.183,\\
m_{int,present,3} &\approx& 0.312.
\end{eqnarray}

Therefore, we can observe from the results of Eqs. (\ref{maria111}), (\ref{maria222}), (\ref{maria333}) and (\ref{maria444}) that the presence of the interaction between the Dark Sectors affects the present day values of the cosmographic parameters, in particular they are lower than the case without interaction.\\
Also for the interacting case, we obtain values of the cosmographic parameters we have considered which are in agreement (or within the errors) with the results obtained in some recent papers.\\

\section{Squared Speed Of The Sound}
We now consider and study an important quantity considered in Cosmology in order to check the stability of any DE model taken into account: this quantity is known as squared speed of sound, it is indicated with $v_s^2$ and it is generally defined as follows \cite{myung}:
\begin{eqnarray}
v_s^2 = \frac{\dot{p}_{tot}} {\dot{\rho}_{tot}}  = \frac{\dot{p}_D}{\dot{\rho}_D + \dot{\rho}_m} , \label{genvs}
\end{eqnarray}
where $p_{tot}= p_D$ and $\rho_{tot} = \rho_D + \rho_m$ are, respectively, the total pressure and the total energy density of the
DE model taken into account.\\
The sign of the squared speed of the sound $v_{s}^{2}$ assumes an important role if we want to study the stability of a background
evolution. In fact, a negative value of the squared speed of the sound $v_s^2$ indicates a classical instability of a given perturbation in General Relativity \cite{myung,kim}. In the paper of Myung \cite{myung}, it was observed that the sing of the squared speed of the sound $v_{s}^{2}$ for the HDE model remains always negative if the future event horizon is considered as IR cut-off (which means this model is unstable), while for the Chaplygin gas and the tachyon it is observed to be positive defined. Moreover, Kim et al. \cite{kim} obtained that $v_{s}^{2}$ for the Agegraphic DE (ADE) model is always negative, which leads to an instability of the perfect fluid for the model. Recently, Sharif $\&$ Jawad \cite{sharif} have shown that the interacting new HDE model is characterized by a negative  squared speed of the sound $v_{s}^{2}$.  Jawad et al. \cite{jawad} have shown that the $f\left(G\right)$ modified gravity model in the HDE scenario with the choice of the scale factor in power law form is classically unstable. Pasqua et al. \cite{miovs2} showed  that the DE model  based on the Generalized Uncertainty Principle (GUP) with power-law form of the scale factor $a\left(t\right)$ is classically instable.\\
We now want to study the behavior of the squared speed  $v_s^2$ for the model we consider in this paper for both non interacting and later on interacting Dark Sectors.

\subsection{Non Interacting Case}
\begin{figure}[ht]
\centering\includegraphics[width=8cm]{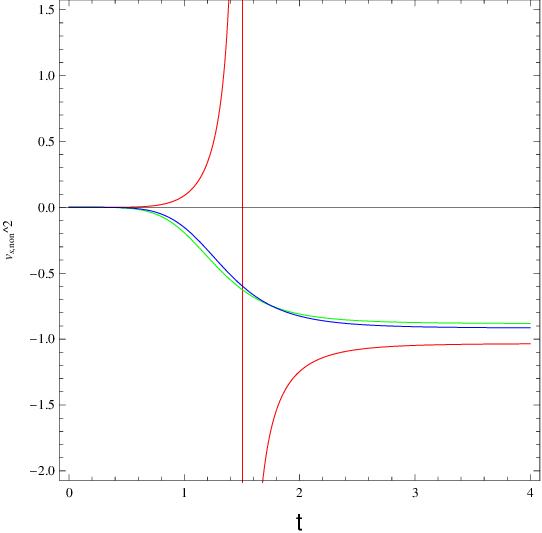}
\caption{Plot of the squared speed of the sound $v_{s,non}^2$ given in Eq. (\ref{vsnonlulu}) for the non interacting case. } \label{vsnontoto}
\end{figure}

\begin{figure}[ht]
\centering\includegraphics[width=8cm]{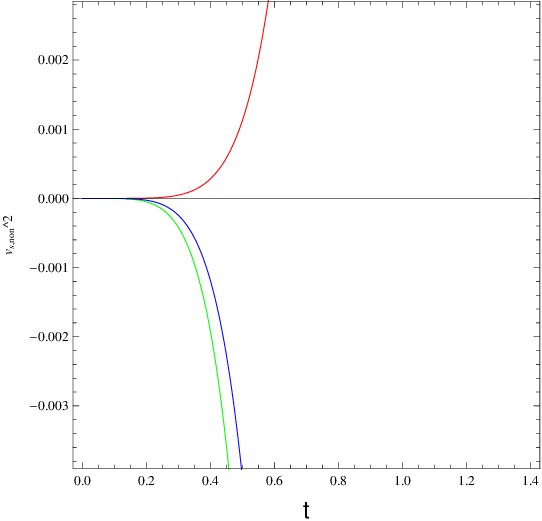}
\caption{Plot of the squared speed of the sound $v_{s,non}^2$ given in Eq. (\ref{vsnonlulu}) for the non interacting case for $0<t<1.4$. } \label{vsnontoto2}
\end{figure}

We start studying the behavior of $v_s^2$ for  the non interacting case.\\
Using in Eq. (\ref{genvs}) the expressions of the pressure of DE $p_D$, the energy density of $\rho_D$ and  the energy density of DM $\rho_m$ given, respectively, in Eqs. (\ref{rhod2}) , (\ref{15}) and (\ref{rhom}), we obtain the following expression for the squared speed of the sound $v_{s,non}^2$ for the non interacting case:
\begin{eqnarray}
v_{s,non}^2&=&\frac{(m-2) (m+3 n-2)  \left[2 \alpha +n (n \beta -\epsilon )\right] \phi _0t^{m-3}}{3n \left\{\frac{ n t^{-1-3 n} \rho_{m0}}{a_0^3}-
(m-2)  \left[2 \alpha +n (n \beta -\epsilon )\right] \phi_0t^{m-3}\right\}} \nonumber \\
&=& \frac{(m-2) (m+3 n-2)  \left[2 \alpha +n (n \beta -\epsilon )\right] \phi _0}{3n \left\{\frac{ n t^{2-3 n-m} \rho_{m0}}{a_0^3}-
(m-2)  \left[2 \alpha +n (n \beta -\epsilon )\right] \phi_0\right\}} . \label{vsnonlulu}
\end{eqnarray}
In Figures  \ref{vsnontoto} and \ref{vsnontoto2},  we plot the behavior of the squared speed of the sound $v_{s,non}^2$ obtained in Eq. (\ref{vsnonlulu}) for the range of values $0<t<4$ and $0<t<1.4$, respectively.
We can observe in Figures \ref{vsnontoto} and \ref{vsnontoto2} that the model corresponding to $m=2.2$ is initially stable since we obtain $v_s^2\geq 0$, but it changes its sign for $t>1.5$ becoming unstable; moreover, for late times, it assumes a constant value. Instead, the models corresponding to $m=1.5$ and $m=1.3$ are unstable for all the range of values of $t$ since we obtain $v_s^2\leq 0$; moreover, these two models have a smooth decreasing behavior,  tending to constant values for late times (i.e. for $t\rightarrow \infty$).\\
We can now make some considerations about the expression of $v_s^2$ obtained in Eq. (\ref{vsnonlulu}). \\
For $t\rightarrow 0$, we have that $v_s^2\rightarrow 0$, as we can also observe from Figures \ref{vsnontoto} and \ref{vsnontoto2}.\\
For far future, i.e. for $t\rightarrow \infty$, we obtain that $v_s^2$ assumes a constant value given by the following general relation:
\begin{eqnarray}
v_{s,non,late}^2=\frac{2-m-3 n}{3 n }. \label{vsnon}
\end{eqnarray}
Therefore, for the three different cases considered in this paper, we obtain:
\begin{eqnarray}
v_{s,non,late,1}^2 &\approx& -1.033, \label{}\\
v_{s,non,late,2}^2&\approx&-0.917,  \label{}\\
v_{s,non,late,3}^2&\approx&-0.883. \label{}
\end{eqnarray}
The value $v_{s,non,late}^2=0$ (which indicates transition from a stable to an unstable model) can be obtained for the values given by the  combination of $m$ and $n$ $m+3n-2=0$, which implies the following relation between $m$ and $n$:
\begin{eqnarray}
m = 2-3n. \label{mn}
\end{eqnarray}
In order to have a stable model, the following conditions must be fulfilled:
\begin{eqnarray}
2-m -3n&>&0, \label{}\\
n&>&0. \label{}
\end{eqnarray}
We know that the second condition is always satisfied since we want to have an accelerated Universe.
Instead, in order to have an unstable model, the following conditions must be fulfilled:
\begin{eqnarray}
2-m -3n&<&0, \label{}\\
n&>&0. \label{}
\end{eqnarray}
We know that the second condition is always satisfied since we want to have an accelerated Universe.\\
At present time, for the three different cases we are taking into account, we have that the squared speed of the sound assumes the following values:
\begin{eqnarray}
v_{s,non,present,1}^2 &\approx& 0.065,\\
v_{s,non,present,2}^2 &\approx& -0.122,\\
v_{s,non,present,3}^2 &\approx& -0.158.
\end{eqnarray}
Therefore, we can conclude that, at present time, for the first case considered, we deal with a stable model since $v_s^2$ assumes a positive value while for the other two cases considered we deal with an unstable model since $v_s^2$ assumes negative values.

\subsection{Interacting Case}
\begin{figure}[ht]
\centering\includegraphics[width=8cm]{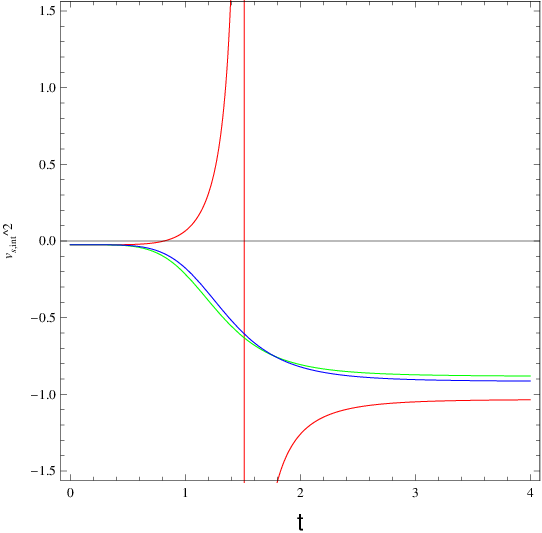}
\caption{Plot of the squared speed of the sound  $v_{s,int}^2$ given in Eq. (\ref{vsintlu}) for the interacting case. } \label{vsint}
\end{figure}

\begin{figure}[ht]
\centering\includegraphics[width=8cm]{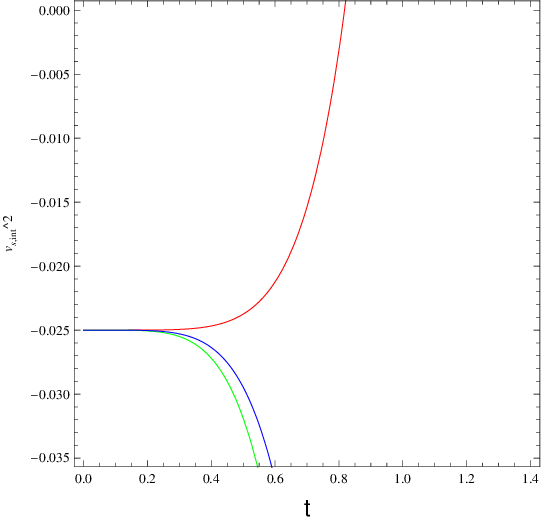}
\caption{Plot of the squared speed of the sound  $v_{s,int}^2$ given in Eq. (\ref{vsintlu}) for the interacting case for $0<t<1.4$. } \label{vsint2}
\end{figure}

We now want to study the behavior of the squared speed of the sound $v_s^2$ for the interacting case.\\
Using in the general expression of $v_s^2$ given in Eq. (\ref{genvs}) the expressions of $\rho_D$, $p_D$ and $\rho_m$ for the interacting case obtained, respectively, in Eqs. (\ref{rhod2}),  (\ref{rhomint}) and (\ref{14-2}) we obtain the following relation for $v_s^2$ for the interacting case:
\begin{eqnarray}
v_{s,int}^2=-\frac{\frac{3 b^2 \left(-1+b^2\right) n t^{-1+3 \left(-1+b^2\right) n} \rho _{m0}} {a_0^3}+\frac{(m-2) (m+3 n-2) t^{-3+m} \left[2\alpha +n (n \beta -\epsilon )\right] \phi _0}{n}} {\frac{3 \left(-1+b^2\right) n t^{-1+3 \left(-1+b^2\right) n} \rho _{m0}} {a_0^3}+3 (m-2) t^{-3+m}\left[2 \alpha +n (n \beta -\epsilon )\right] \phi _0}. \label{vsintlu}
\end{eqnarray}
In the limiting case of $b^2=0$, we recover the same result of the non interacting case obtained in the previous subsection.\\
In Figures \ref{vsint} and \ref{vsint2},  we plot the behavior of the squared speed of the sound $v_{s,int}^2$ given in Eq. (\ref{vsintlu}) for  for the range of values $0<t<4$ and $0<t<1.4$, respectively. \\
In Figure \ref{vsint2}, it is evident the effect of the interaction on the initial value of the squared speed of the sound. In particular, we have that the presence of interaction between the two Dark Sectors leads to lower present day value of the squared speed of the sound $v_s^2$ if compared with the non interacting case we studied in the previous subsection. \\
We can observe in Figures \ref{vsint} and \ref{vsint2} that the model corresponding to $m=2.2$ is initially unstable since we obtain $v_s^2<0$ but it becomes stable for $t\geq 0.6$. Moreover, for $t>1.5$, the model becomes unstable again. Instead, the models corresponding to the cases with $m=1.5$ and $m=1.3$ are unstable for all the range of values of $t$ since we obtain $v_s^2<0$; moreover, the cases corresponding to $m=1.5$ and $m=1.3$ have a smooth decreasing behavior and they assume a constant value for late times. We can also conclude that all the three cases considered lead to an unstable model at present time, i.e. at $t=0$.\\
In the limiting case corresponding to $m=2$, we obtain that $v_s^2 = -b^2$.\\
For $t\rightarrow 0$, we obtain that $v_s^2 = -b^2 = -0.025$, as we can also observe in Figure \ref{vsint2}.\\
In the limiting case of $t\rightarrow \infty$, we obtain the same results of the non interacting case obtained in the previous subsection.\\
At present time, we have, for the three cases considered, that:
\begin{eqnarray}
v_{s,int,present,1}^2 &\approx& 0.041,\\
v_{s,int,present,2}^2 &\approx& -0.147,\\
v_{s,int,present,3}^2 &\approx& -0.182.
\end{eqnarray}
As for the non interacting scenario, we obtain that  for the first case considered, we deal with a stable model since $v_s^2$ assumes a positive value while for the other two cases considered we deal with an unstable model since $v_s^2$ assumes negative values. Moreover , we conclude that the presence of interaction affects the behavior of the squared speed of the sound $v_s^2$. In  particular, we have that the value for $t\rightarrow 0$ is lower than the value obtained for the non interacting case.

\section{Conclusions}
In this paper, we considered a recently proposed  DE energy density model $\rho_D$ proposed by Chen $\&$ Jing \cite{modelhigher} which is function of the Hubble parameter $H$ and of the first and the second time derivatives of $H$ in the framework of the Chameleon Brans-Dicke (BD) Cosmology. Moreover, this DE model is also characterized by three free parameters indicated with $\alpha$, $\beta$ and $\varepsilon$. We have considered a particular ansatz for the scale factor $a$,  the BD scalar field $\phi$,  $V$  and $f$. In particular, we have chosen $a$ and $\phi$ as power law form of the time, while $V$ and $f$ are taken as power laws of $\phi$ following the same choice made in \cite{33}. The main purpose of this paper is to study the cosmological consequences of considering the said DE model in interacting as well as non-interacting scenario in the framework of Chameleon Brans-Dicke (BD) Cosmology. For the interacting scenario the interaction term is taken as $Q=3b^2H\rho_m$, with $b^2$ indicating the interaction strength. In this two-component model, the matter sector is chosen to be pressureless DM. As for any discussion related to any DE model, the EoS parameter plays a  significant role when the behavior of a particular model is wanted to be studied. For this reason, we have studied the behaviour of the reconstructed Effective EoS parameter in the framework of Chameleon BD gravity. The plot of the effective EoS parameter $w_{eff}$ we derived in Eq. (\ref{omegaeff}) against the cosmic time $t$ for the non interacting case (see Figure \ref{omegaeffnon}) shows that, initially, the effective EoS parameter $w_{eff}$ is much greater that $-1$. However, as the cosmic time $t$ passes, we observe that $w_{eff}$ starts to behave like a monotone decreasing function of the cosmic time $t$. We can also observe that, in the later stage, $w_{eff}$ is tending  to the value $-1$ and, at $t\approx 1.4$, it is crossing the phantom boundary of $-1$ for the cases corresponding to $m=2.2$. Instead, for the other two values of $m$ we considered, i.e. $m=1.5$ and $m=1.3$, the effective EoS parameter $w_{eff}$ is getting asymptotically near the phantom boundary without crossing it. Hence, in the non-interacting scenario, we conclude that the effective EoS parameter $w_{eff}$ has a``quintom"-like behavior for the cases corresponding to $m=2.2$. A similar behaviour of the effective EoS parameter $w_{eff}$ is observed in Figure \ref{omegaeffint} for the interacting case. Thus, the presence of the interaction between the two Dark Sectors (which is characterized by the expression of $Q$ we have chosen) does not bring any significant change in the behavior of the effective EoS parameter $w_{eff}$. \\
Apart from $w_{eff}$, we have also studied the behavior of the EoS parameter of DE $w_D$ for both non interacting and later on interacting Dark Sectors. For the non interacting case, we observed that it can assume values which can be greater, equals or less than $-1$ (i.e. the value corresponding to the $\Lambda$CDM model) according to the values assumed by the parameters involved: in particular, an important role on the final value of $w_D$ is assumed by the two parameters $m$ and $n$. In particular, we observed that $w_D>-1$ (quintessence) for $m>2$ while we have that $w_D<-1$ (phantom) when $m<2$, where $n$ is kept at positive level in all cases (see Figure  \ref{omegacham}). For the interacting case, we obtained the following present day values of the EoS parameter of DE: $w_{D,int,present,1}   \approx -1.047$, $w_{D,int,present,2}   \approx -0.930$ and $w_{D,int,present,3}   \approx -0.897$. Therefore, for the case with $m=2.2$, we obtain a value which is beyond the phantom divide line, while for the other two cases we obtain values higher than -1.\\
Moreover, for the first two cases considered, i.e. for $m=2.2$ and $m=1.5$, we have that the value of $w_{D,int}$ we obtained lie within the constraints obtained through observations given in Table 1.\\
We have also studied the behavior of the deceleration parameter $q$, which is plotted against the cosmic time $t$ for the non-interacting scenario in Figure \ref{qnon} and for the interacting scenario in Figure \ref{qint}. The transition from decelerated to accelerated phase of the Universe (i.e. the transition from $q>0$ to $q<0$) is apparent at $t\approx 0.8$ for both non interacting and interacting DE and DM and for all the three cases we considered. Therefore, irrespective of presence of interaction term, the transition from decelerated to accelerated phase of Universe is attainable. Moreover, for both non interacting and interacting scenarios and for all the three cases we considered, we obtained negative values of the present day  deceleration parameter $q$, which indicated an accelerated Universe, in agreement with the most recent cosmological data available.  \\
In order to have further information about the model we are studying, we have studied and plotted the statefinder trajectories for both the non interacting and the interacting scenarios.
The statefinder trajectories for the non interacting scenario are plotted in Figure \ref{statenon}.  For the three cases considered in this paper, we obtained the following values of the present day values of statefinder pair: $\left\{r_1,s_1   \right\}_{non,present} \approx  \left\{ 1.099, -0.033  \right\}$, $\left\{r_2,s_2   \right\}_{non,present}  \approx  \left\{ 0.777,0.083   \right\}$ and $\left\{r_3,s_3   \right\}_{non,present}  \approx  \left\{  0.698, 0.117 \right\}$.  Therefore, at present time, we obtain that for the first case considered, i.e. for $m=2.2$, the values of the statefinder parameters  slightly differ from the value corresponding to the $\Lambda$CDM model. Instead, for the cases with $m=1.5$ and $m=1.3$, we obtain values of the statefinder parameters which considerably differs from that of the $\Lambda$CDM. Moreover, for the case corresponding to $m=2.2$, we obtained a quintessence-like model  while for the other two cases considered we obtain a phantom-like model.\\
The statefinder trajectories for the interacting scenario are plotted in Figure \ref{stateint}. We observed that the $\Lambda$CDM point is attainable for the case corresponding to $m=2.2$. Furthermore, the trajectory can go beyond that fixed point $\{r=1,s=0\}$ and it can reach the quadrant $r>1,s<0$. Instead, for the cases with $m=1.3$ and $1.5$, the $\Lambda$CDM is not attainable by the model. The present day values of the statefinder pair for the interacting scenario are given, for the three different cases we consider in this paper,  by $\left\{r_{1}, s_{1} \right\}_{int,present}\approx \left\{1.060, -0.020\right\}$, $\left\{r_{2},s_{2}  \right\}_{int,present} \approx \left\{0.738,0.096\right\}$ and $\left\{r_{3}, s_{3} \right\}_{int,present}\approx \left\{0.659,  0.129\right\}$. Therefore, at present time, we obtain that for the first case considered, i.e. for $m=2.2$, the values of the statefinder parameters  slightly differ from the value corresponding to the $\Lambda$CDM model. Instead, for the cases with $m=1.5$ and $m=1.3$, we obtain values of the statefinder parameters which considerably differs from that of the $\Lambda$CDM. Moreover, for the case corresponding to $m=2.2$, we obtained a quintessence-like model  while for the other two cases considered we obtain a phantom-like model.  Finally, we can conclude that under interacting scenario, the current Universe does not favour the $\Lambda$CDM scenario under this model. \\
We also studied the behavior of some cosmographic parameters, in particular the jerk $j$, the snap $s_{cosmo}$, the lerk $l$ and the work-out $m$ parameters for the model we are dealing with. We have also calculated the present day values of all the cosmographic parameters considered. For both non interacting and interacting scenario, we obtained the present day values of the four cosmographic parameters considered which are in agreement with results found in recent papers.\\
Finally, we studied the behavior of an important quantity in Cosmology, i.e. the squared speed of the sound $v_s^2$, for the model we are studying for both non interacting and interacting scenarios. The squared speed of the sound is used in order to understand if the model we study is stable or unstable against small perturbations. We observed that the model can be both stable or unstable depending on the values of the parameters we considered. In particular, for the non interacting scenario, we obtained that, for the case with $m=2.2$, $v_s^2$ is initially stable but it becomes unstable for later times. For the other two cases, i.e. for $m=1.5$ and $m=1.3$, we obtain that $v_s^2$ is always unstable, becoming a constant value for late times. The present day values of $v_s^2$ for the non interacting scenario for the three different cases we are taking into account are given, respectively, by: $v_{s,non,present,1}^2 \approx 0.065$, $v_{s,non,present,2}^2 \approx -0.122$ and $v_{s,non,present,3}^2 \approx -0.158$. Then, at present time, for the first case considered, we obtain an stable model since $v_s^2$ assumes a positive value while for the other two cases considered we deal with an unstable model since $v_s^2$ assumes negative values. For the interacting scenarios, we have that, for $m=2.2$, $v_s^2$ is initially unstable, then it becomes stable for a short period of time and then it becomes unstable again. Instead, for the other two cases considered, i.e. for $m=1.5$ and $m=1.3$, we obtain that the squared speed of the sound $v_s^2$ is always unstable, becoming a constant value for late times. At present time, we obtain the following values for $v_s^2$ for the interacting scenario:  $v_{s,int,present,1}^2 \approx 0.0410$, $v_{s,int,present,2}^2 \approx -0.147$ and  $v_{s,int,present,3}^2 \approx -0.182$.
As for the non interacting scenario, we obtain that  for the first case considered, we deal with a stable model since $v_s^2$ assumes a positive value while for the other two cases considered we deal with an unstable model since $v_s^2$ assumes negative values. Moreover, we conclude that the presence of interaction affects the values of the present day  squared speed of the sound $v_s^2$ since the interaction term affect in particular the values of $v_s^2$ for $t\rightarrow 0$ and the present day values we derived.

\section{Conflict of interest}
The authors declare that they have no conflict of interest.

\section{Acknowledgement}
S. Chattopadhyay acknowledges financial support from the Council
of Scientific and Industrial Research (Govt of India) with Grant
No. 03(1420)/18/EMR-II. The authors express their sincere thanks
to Dr. Ines G. Salako for some constructive discussions while
carrying out the work.

\section{Data Availability statement}
The authors hereby declare that no primary data have been used in
this work. The cosmological parameters, as and when required, have
been compared to the outcomes of experiments available in the
existing standard literatures that are appropriately cited.


\begin{thebibliography} {99}
\bibitem{odi1} S. Nojiri, S. D. Odintsov, Gen Relativ Gravit \textbf{38}, 1285
(2006)
\bibitem{1b}  A. G. Riess et al.,  Astron. J. \textbf{116},  1009 (1998)
\bibitem{1a}  S. Perlmutter et al., Astrophys. J. \textbf{517},  565 (1999)
\bibitem{1c}  U. Seljak, et al., Phys. Rev. D \textbf{71},  103515 (2005)
\bibitem{1d}  P. Astier et al., Astron. Astrophys. \textbf{447},  31 (2006)
\bibitem{1h}  K. Abazajian  et al., Astron. J. \textbf{129},   1755 (2005)
\bibitem{cmb2}  D.N. Spergel et al., Astrophys. J. Suppl. Ser. \textbf{148},   175 (2003)
\bibitem{cmb3}  E. Komatsu et al., Astrophys. J. Suppl. \textbf{180},   330 (2009)
\bibitem{planck}  Planck Collaboration, P.A.R. Ade et al., \ 2013, arXiv:1303.5076
\bibitem{sds1}  M. Tegmark et al., Phys. Rev. D \textbf{69},   103501 (2004)
\bibitem{sds2}  K. Abazajian  et al., Astron. J. \textbf{128},   502 (2004)
\bibitem{xray}  S.W. Allen et al., Mon. Not. Roy. Astron. Soc. \textbf{353},   457 (2004)
\bibitem{copeland-2006}  E.J. Copeland, M. Sami,  M. Tsujikawa,  International Journal of Modern Physics D \textbf{15},   1753  (2006)
\bibitem{delcampo}  S. del Campo, R. Herrera, D. Pavon, J. Cosmol. Astropart. Phys. \textbf{0901},   020 (2009)
\bibitem{delcampoa}  G. Leon, E.N. Saridakis, Phys. Lett. B \textbf{693},   1 (2010)
\bibitem{delcampob}  J.B. Jimenez, A.L. Maroto, AIP Conf. Proc. \textbf{1122},   107 (2009)
\bibitem{delcampoc}  M.S. Berger, H. Shojae, Phys. Rev. D \textbf{73},   083528 (2006)
\bibitem{delcampod}  X. Zhang, Mod. Phys. Lett. A \textbf{20},   2575 (2005)
\bibitem{delcampoe}  K. Griest, Phys. Rev. D \textbf{66},   123501 (2002)
\bibitem{delcampof}  M. Jamil, F. Rahaman,  Eur. Phys. J. C \textbf{64},   97 (2009)
\bibitem{delcampog}  M. Jamil, E.N. Saridakis, M.R. Setare, Phys. Rev. D \textbf{81},   023007 (2010)
\bibitem{delcampoh}  M. Jamil,  E.N. Saridakis, J. Cosmol. Astropart. Phys. \textbf{07},   028 (2010)
\bibitem{twothirds}  H.V. Peiris et al., Astrophys. J. Suppl. Ser. \textbf{148},   213 (2003)
\bibitem{dil1}  N. Arkani-Hamed, P. Creminelli, S. Mukohyama, M. Zaldarriaga, J. Cosmol. Astropart. Phys. \textbf{4},   1 (2004)
\bibitem{dil2}  M. Gasperini, F. Piazza, G. Veneziano, Phys. Rev. D \textbf{65},   023508 (2002)
\bibitem{dil2-1}  E. Elizalde et al., Euro. Phys. J. C \textbf{53},   447 (2008)
\bibitem{kess1}  C. Armendariz-Picon, C., Mukhanov, V., Steinhardt, P. J., Phys. Rev. D \textbf{63},   103510 (2001)
\bibitem{kess3}  T. Chiba, T. Okabe, M. Yamaguchi,  Phys. Rev. D \textbf{62},   023511 (2000)
\bibitem{kess4}  C.A. Picon, T. Damour,  V.  Mukhanov,  Phys. Lett. B \textbf{458},   209 (1999)
\bibitem{kess2-2}  A. Sen,  Modern Physics Letters A \textbf{17},   1797 (2002)
\bibitem{quint1} B. Ratra, P.J.E.   Peebles,  Phys. Rev. D \textbf{37},   3406 (1988)
\bibitem{quint3}  I. Zlatev, L. Wang, P.J.   Steinhardt, Physical Review Letters \textbf{82},   896 (1999)
\bibitem{quint5}  C. Wetterich, Nucl. Phys. B \textbf{302},   668 (1988)
\bibitem{quint6}  Caldwell, R. R., Dave, R.,   Steinhardt, P. J., Phys. Rev. Lett. \textbf{80},   1582 (1998)
\bibitem{quint7}  M. Doran, J.   Jaeckel, Phys. Rev. D \textbf{66},   043519 (2002)
\bibitem{tac3}  T. Padmanabhan, T.R. Choudhury,  Phys. Rev. D \textbf{66},   081301 (2002)
\bibitem{tac1-1}  J.S. Bagla, H.K. Jassal, T. Padmanabhan,  Phys. Rev. D \textbf{67},   063504 (2003)
\bibitem{tac1-2}  Y. Shao, Y.X. Gui, W. Wang,  Modern Physics Letters A \textbf{22},   1175 (2007)
\bibitem{tac1-3}  G. Calcagni, A.R. Liddle, Phys. Rev. D \textbf{74},   043528 (2006)
\bibitem{tac1-4}  E.J. Copeland et al., Phys. Rev. D \textbf{71},   043003 (2005)
\bibitem{tac2-1}  A. Sen, Journal of High Energy Physics \textbf{10},   8 (1999)
\bibitem{tac2-3}  E.A. Bergshoeff  et al., Journal of High Energy Physics \textbf{5},   9 (2000)
\bibitem{tac2-4}  J. Kluson,  Phys. Rev. D \textbf{62},   126003 (2000)
\bibitem{pha1} R.R. Caldwell,  Phys. Lett. B \textbf{545},   23 (2002)
\bibitem{pha3}  S. Nojiri,  S.D. Odintsov,  Physics Letters B \textbf{562},   147 (2003)
\bibitem{pha4}  B. McInnes,  J. High Energy Phys. \textbf{0208},   029 (2002)
\bibitem{pha5}  L.P. Chimento, R.  Lazkoz, Phys. Rev. Lett. \textbf{91},   211301 (2003)
\bibitem{pha6}  B. Boisseau, G.  Esposito-Farese, D. Polarski, A.A. Starobinsky,  Phys. Rev. Lett. \textbf{85},   2236 (2000)
\bibitem{pha7}  R. Gannouji, D. Polarski, A. Ranquet, A.A. Starobinsky,  J. Cosmol. Astropart. Phys. \textbf{0609},   016 (2006)
\bibitem{qui1} A. Anisimov, J. Cosmol. Astropart. Phys. \textbf{6},   6 (2005)
\bibitem{qui2} E. Elizalde,  S. Nojiri, S.D. Odintsov, Phys. Rev. D \textbf{70},   043539 (2004)
\bibitem{qui3} S. Nojiri, S.D. Odintsov, S. Tsujikawa, Phys. Rev. D \textbf{71},   063004 (2005)
\bibitem{qui4}  B. Feng, X.L. Wang, X.M.  Zhang, Phys. Lett. B \textbf{607},   35 (2005)
\bibitem{qui5} Z.K. Guo, Y.S. Piao, X.M. Zhang, Y.Z.   Zhang,  Phys. Lett. B \textbf{608},   177 (2005)
\bibitem{qui6}  Y.F. Cai, M.Z.  Li, J.X. Lu, Y.S. Piao, T.T. Qiu, X. M.    Zhang,  Phys. Lett. B \textbf{651},   1 (2007)
\bibitem{qui8}  W. Zhao, Y.   Zhang,  Phys. Rev. D \textbf{73},   123509 (2006)
\bibitem{qui10}  H. Mohseni Sadjadi, M.   Alimohammadi, Phys. Rev. D \textbf{74},   043506 (2006)
\bibitem{qui12}  M.R. Setare,    E. N. Saridakis,  J. Cosmol. Astropart. Phys. \textbf{0809},   026 (2008)
\bibitem{cgas1}  A. Kamenshchik, U. Moschella, V. Pasquier,  Physics Letters B \textbf{511},  265 (2001)
\bibitem{cgas2}  M.C. Bento, O. Bertolami, A.A.  Sen,  Phys. Rev. D \textbf{66},   043507 (2002)
\bibitem{cgas3}  M.R. Setare, European Physical Journal C \textbf{52},   689 (2007)
\bibitem{ade1}  H. Wei, R.G.  Cai,   Physics Letters B \textbf{660},   113 (2008)
\bibitem{ade2}  R.G. Cai,  Phys. Lett. B \textbf{657},   228 (2007)
\bibitem{2} T.  Padmanabhan,   Phys. Rep. \textbf{380},   235, (2003)
\bibitem{2a} Y.F. Cai,   E. N. Saridakis,  M.R. Setare, J.Q.  Xia,   Phys. Rep. \textbf{493},   1 (2010)
\bibitem{3}  M. Li,  Physics Letters B \textbf{603},   1 (2004)
\bibitem{3b}  Y.S. Myung, M.G.   Seo,  Physics Letters B \textbf{671},   435 (2009)
\bibitem{4} Q.G. Huang, M.    Li,  J. Cosmol. Astropart. Phys. \textbf{8},   13 (2004)
\bibitem{5}  G. 't Hooft,   International Journal of Modern Physics D \textbf{15},   1587 (2006)
\bibitem{5a}  L. Susskind,  Journal of Mathematical Physics \textbf{36},  6377 (1995)
\bibitem{5b}  D. Bigatti,    L. Susskind,   Strings, Branes and Gravity  TASI \textbf{99},   883 (2001)
\bibitem{6}  W. Fischler,    L. Susskind,  1998, arXiv,hep-th/9806039
\bibitem{7}  A. G. Cohen, D.B. Kaplan, A.E.   Nelson,  Physical Review Letters \textbf{82},   4971 (1999)
\bibitem{n2primo}  M. Li, X.D.  Li, S.  Wang, Y. Wang  X. Zhang,  J. Cosmol. Astropart. Phys.  \textbf{0912},   014 (2009)
\bibitem{n2secondo}  M. Li, X.D.  Li, S.  Wang, X. Zhang,  J. Cosmol. Astropart. Phys.  \textbf{0906},   036 (2009)
\bibitem{nojo}  S. Nojiri,  S.D.  Odintsov,  2008, arXiv:0807.0685
\bibitem{nojo2}  S. Nojiri,  S.D.  Odintsov, Int. J. Geom. Meth. Mod. Phys. \textbf{4},   115 (2007)
\bibitem{frieman1} J.A. Frieman, M.S. Turner, D. Huterer,  Annu. Rev. Astro. Astrophys. \textbf{46}, 385 (2008)
\bibitem{frieman2} S. Nojiri, S.D.  Odintsov,  Phys. Lett. B, \textbf{19}, 627 (2004)
\bibitem{15}  C.J. Fen,  X.Z. Li, Physics Letters B \textbf{679},   151 (2009)
\bibitem{15a}  C.J. Feng,   X. Zhang, Physics Letters B \textbf{680},   399 (2009)
\bibitem{15b}  H. Wei, Nuclear Physics B \textbf{819},   210 (2009)
\bibitem{15c}  Y. Bisabr, General Relativity and Gravitation \textbf{41},   305 (2009)
\bibitem{15e}  K. Nozari, N.  Rashidi, International Journal of Theoretical Physics \textbf{48},   2800 (2009)
\bibitem{15g}  K. Karami,  M.S. Khaledian, Journal of High Energy Physics \textbf{3},   86 (2011)
\bibitem{15i} M.R. Setare,    M. Jamil,  EPL (Europhysics Letters) \textbf{92},   49003 (2010)
\bibitem{15l} M.R. Setare,    M. Jamil, Physics Letters B \textbf{690},   1 (2010)
\bibitem{mio}  A. Pasqua,    I. Khomenko,  International Journal of Theoretical Physics \textbf{52}, 3981 (2013)
\bibitem{bra1}  C. Deffayet, G. Dvali, G. Gabadadze, Phys. Rev. D \textbf{65},   044023 (2002)
\bibitem{bra2}  V. Sahni, Y. Shtanov,  J. Cosmol. Astropart. Phys. \textbf{11},   14 (2003)
\bibitem{miors}  J. Ovalle,  F. Linares, A. Pasqua,   A. Sotomayor,  Classical and Quantum Gravity \textbf{30},  175019 (2013)
\bibitem{antonico1} A. Pasqua, S. Chattopadhyay, R. Myrzakulov, 2015, arXiv:1511.00611
\bibitem{antonico2} A. Pasqua, S. Chattopadhyay,  M. Khurshudyan, et al.,  International Journal of Theoretical Physics \textbf{54}, 972 (2015)
\bibitem{antonico3} A. Jawad,  S. Chattopadhyay,  A. Pasqua, European Physical Journal Plus \textbf{129}, 51 (2014)
\bibitem{mioft1}  A. Pasqua,   S. Chattopadhyay,  Canadian Journal of Physics \textbf{91},   351 (2013)
\bibitem{mioft2}  S. Chattopadhyay,    A. Pasqua, Astrophys. Space Sci. \textbf{344},   269 (2013)
\bibitem{mioft3}  R. Ghosh, A. Pasqua,   S. Chattopadhyay,  European Physical Journal Plus \textbf{128},   12 (2013)
\bibitem{ft1}  G.R. Bengochea, R. Ferraro,  Phys. Rev. D \textbf{79},   124019 (2009)
\bibitem{ft2}  E.V. Linder,  Phys. Rev. D \textbf{81},   127301 (2010) [Erratum-ibid. D \textbf{82},   109902 (2010)]
\bibitem{ft3}  B. Li, T.P.Sotiriou, J.D. Barrow,   Phys. Rev. D \textbf{83},   064035 (2011)
\bibitem{ft5} M. Li, R.X.  Miao, Y.G. Miao,  JHEP \textbf{1107},   108 (2011)
\bibitem{ft6}  Karami, K.,  Abdolmaleki, A.,  J. Cosmol. Astropart. Phys. \textbf{1204},   007 (2012)
\bibitem{ft8}  K. Bamba, C.Q. Geng, C.C. Lee, L.W. Luo,  J. Cosmol. Astropart. Phys. \textbf{1101},   021 (2011)
\bibitem{fr1}  S. Capozziello,   Int. J. Mod. Phys. D \textbf{11},   483 (2002)
\bibitem{fr2}  T.P.Sotiriou, V. Faraoni,  arXiv:0805.1726
\bibitem{fr3}  A.A .Starobinsky,  Phys. Lett. B \textbf{91},   99 (1980)
\bibitem{miofr}  A. Jawad,  S. Chattopadhyay,    A. Pasqua, Astrophys. Space Sci. \textbf{346},   273 (2013)
\bibitem{fr7}  S. Nojiri,    S.D.  Odintsov,  Gen. Rel. Grav. \textbf{36},   1765 (2004)
\bibitem{fr8}  N. Arkani-Hamed, H.C. Cheng, M.A.  Luty, S.   Mukohyama,   JHEP \textbf{05},   043528 (2004)
\bibitem{fr10}  S. Nojiri,     S.D. Odintsov,  Phys. Rev. D \textbf{68},   123512 (2003)
\bibitem{fr11} M.C.B.  Abdalla,  S.D. Odintsov,   Class. Qunt. Grav. \textbf{22},   L35 (2005)
\bibitem{fr15}  A. Aghmohammadi, K. Saaidi, M.R.    Abolhassani,  Int. J. Theor Phys. \textbf{49},   709 (2010)
\bibitem{fr12}  S. Capozziello,  S. Nojiri,  S.D.  Odintsov, A.  Troisi,  Phys. Lett. B \textbf{639},   135 (2006)
\bibitem{fr13}  S.A. Appleby, R.A.    Battye,   Phy. Lett. B \textbf{654},   7 (2007)
\bibitem{fr14}  D.A. Easson,  Int. Mod. Phys. A \textbf{19},   5343 (2004)
\bibitem{miofg1}  A. Jawad,  S. Chattopadhyay,    A. Pasqua, European Physical Journal Plus \textbf{128},   88 (2013)
\bibitem{frt2}  R. Myrzakulov, European Physical Journal C \textbf{72},   2203 (2012)
\bibitem{frt5}  F.G.  Alvarenga, A. de la Cruz-Dombriz,  M.J.S.  Houndjo, M.E. Rodrigues,  D.   Saez-Gomez,  Phys. Rev. D \textbf{87},  103526 (2013)
\bibitem{dgp1}  G. Dvali, G. Gabadadze, M.   Porrati, Phys. Lett. B \textbf{485},   208 (2000)
\bibitem{miodbi}  S. Chattopadhyay,    A. Pasqua, International Journal of Theoretical Physics \textbf{52}, 3945 (2013)
\bibitem{miohl}  A. Pasqua,   S. Chattopadhyay,  Astrophys. Space Sci. \textbf{348}, 541 (2013)
\bibitem{miobd2}  A. Pasqua,   S. Chattopadhyay,  Astrophys. Space Sci. \textbf{348}, 283 (2013)
\bibitem{24}  G.S. Greenstein,  Astrophys. Lett. \textbf{1},  139 (1968)
\bibitem{25}  G.S. Greenstein, Astrophys. Space Sci. \textbf{2},  155 (1968)
\bibitem{34} J.  Khoury, A.  Weltman,  Phys. Rev. D \textbf{69},  044026 (2004)
\bibitem{35}  J. Khoury,  A.  Weltman,  Phys. Rev. Lett. \textbf{93},  171104 (2004)
\bibitem{36}   P. Brax, C. van de Bruck, A.C.  Davis,  J. Khoury, A. Weltman,  Phys. Rev. D \textbf{70},  123518 (2004)
\bibitem{37}  H. Farajollahi, A. Salehi,  Astrophys. Space Sci. \textbf{338},  375 (2012)
\bibitem{38}  H. Farajollahi,  A. Ravanpak, G.F.  Fadakar, , Astrophys. Space Sci. \textbf{336},  461 (2011)
\bibitem{33}  Y. Bisabr,  Phys. Rev. D \textbf{86},  127503 (2012)
\bibitem{40}  S. Das, N. Banerjee,  Phys. Rev. D \textbf{78},  043512 (2008)
\bibitem{modelhigher}  S. Chen, J.   Jing,   Physics Letters B \textbf{679},   144 (2009)
\bibitem{goli} L.N. Granda, A. Oliveros,  Physics Letters B \textbf{669}, 275 (2008)
\bibitem{grass1} S. Nojiri,  S.D. Odintsov,  Phys. Rev. D \textbf{72}, 023003 (2005)
\bibitem{grass2} S. Capozziello,  V.F. Cardone, E. Elizalde,  S. Nojiri, S.D.  Odintsov,  Phys. Rev. D \textbf{73}, 043512 (2006)
\bibitem{53}  V. Acquaviva, C. Baccigalupi, S.M. Leach, A.R.  Liddle, F. Perrotta,   Phys. Rev. D \textbf{71},  104025 (2005)
\bibitem{q2}  O. Bertolami, F. Gil Pedro, M.  Le Delliou,     Phys. Lett. B  \textbf{654},  165 (2007)
\bibitem{q2-2} M. Jamil, M.A.  Rashid,   European Physical Journal C \textbf{58}, 111 (2008)
\bibitem{q3} C. Feng, B. Wang, Y. Gong, R.K.  Su,  J. Cosmol. Astropart. Phys. \textbf{9} 5 (2007)
\bibitem{1abd}  E. Abdalla, L.R.  Abramo,  L. Sodr{\'e},  B.  Wang,    Phys. Lett. B  \textbf{673},  107 (2009)
\bibitem{10abd}  O. Bertolami, F. Gil Pedro, M.  Le Delliou,    Gen. Rel. Grav.  \textbf{41},  2839 (2009)
\bibitem{A173}  Z.K. Guo, N. Ohta, S.  Tsujikawa,   Phys. Rev. D  \textbf{76},  023508 (2007)
\bibitem{22abd} J.H. He, B. Wang, P.   Zhang,    Phys. Rev. D  \textbf{80},  063530 (2009)
\bibitem{q1}  L. Amendola, D. Tocchini-Valentini,  Phys. Rev. D  \textbf{64},  043509 (2001)
\bibitem{q1-4}  A. Sheykhi,   M. Jamil,   Phys. Lett. B  \textbf{694},  284 (2011)
\bibitem{q1-8}  W. Zimdahl, D.  Pavon,    Gen. Rel. Grav.  \textbf{35},  413 (2003)
\bibitem{feng08}  C. Feng et al.,   Phys. Lett. B  \textbf{665},  111 (2008)
\bibitem{q4}  K. Ichiki  et al.,   J. Cosmol. Astropart. Phys.  \textbf{06},  005 (2008)
\bibitem{zhang-02-2006}  H. Zhang, Z.H.  Zhu,    Phys. Rev. D  \textbf{73},  043518 (2006)
\bibitem{hins} G. Hinshaw, D. Larson, E. Komatsu, et al., Astrophysical Journal Supplement Series \textbf{208}, 19 (2013)
\bibitem{sah}  V. Sahni, T.D.  Saini, A.A.  Starobinsky, U. Alam,     Soviet Journal of Experimental and Theoretical Physics Letters  \textbf{77},  201 (2003)]\
\bibitem{alam} U.  Alam,  V.  Sahni, T. Deep Saini, A.A.  Starobinsky,     Mon. Not. R. Astron. Soc.   \textbf{344},   1057 (2003)
\bibitem{huang} Z.G.  Huang, X.M. Song, H.Q.  Lu, W. Fang,      Astrophys. Space Sci.  \textbf{315},  175 (2008)
\bibitem{wu1}  P. Wu, H.  Yu,        Physics Letters B  \textbf{693},  415 (2010)
\bibitem{wang}  F.Y. Wang, Z.G.  Dai, S.  Qi,    Astron. Astrophys.  \textbf{507},  53 (2009)
\bibitem{cosmo1} S.  Weinberg,    Gravitation and Cosmology, Wiley.   New York 1972.
\bibitem{cosmo2}  M. Visser,  Class. Quant. Grav.  \textbf{21},  2603 (2004)

\bibitem{capoizzo1} S. Capozziello, L.   Izzo,  Astronomy  \& Astrophysics \textbf{490}, 31 (2008)
\bibitem{capoizzo2} M.V. John, The Astrophysical Journal \textbf{614}, 1 (2004)
\bibitem{capoizzo3} M.V. John, The Astrophysical Journal \textbf{630}, 667 (2008)
\bibitem{orlando2} A. Aviles, C.  Gruber, O.  Luongo, H.   Quevedo, Phys. Rev. D \textbf{86}, 123516 (2012)
\bibitem{arab}  M. Arabsalmani, V.  Sahni,      Phys. Rev. D  \textbf{83},  043501 (2011)
\bibitem{dabro} M.P.   D{\c a}browski,      Physics Letters B  \textbf{625},  184 (2005)
\bibitem{duna} M.  Dunajski, G.   Gibbons,   Classical and Quantum Gravity  \textbf{25}, 235012 (2008)
\bibitem{orlando1} P.K.S. Dunsby, O.   Luongo,  International Journal of Geometric Methods in Modern Physics \textbf{13}, 1630002-606 (2016)
\bibitem{myung}   Y.S. Myung, Phys. Lett. B \textbf{652},  223 (2007)
\bibitem{kim}  K.Y.  Kim, H.W.  Lee, Y.S.  Myung,  Phys. Lett. B \textbf{660},   118 (2008)
\bibitem{sharif}  M. Sharif,  A. Jawad,  Eur. Phys. C \textbf{72},  2097 (2012)
\bibitem{jawad}  A. Jawad,  A. Pasqua, S. Chattopadhyay,  Astrophys. Space Sci. \textbf{344},  489 (2013)
\bibitem{miovs2}  A. Pasqua, S. Chattopadhyay,  I. Khomenko, International Journal of Theoretical Physics \textbf{52}, 2496 (2013)








\end{thebibliography}
\end{document}